\documentclass[12pt,preprint]{aastex}

\slugcomment{}

\shorttitle{Star formation in distant ULIRGs}
\shortauthors{Farrah et al}

\begin{document}

\title{The nature of star formation in distant ultraluminous infrared galaxies selected in a remarkably narrow redshift range}

\author{D. Farrah\altaffilmark{1}}
\author{C. J. Lonsdale\altaffilmark{2,3,4}}
\author{D. W. Weedman\altaffilmark{1}}
\author{H. W. W. Spoon\altaffilmark{1}}
\author{M. Rowan-Robinson\altaffilmark{5}}
\author{M. Polletta\altaffilmark{6}}
\author{S. Oliver\altaffilmark{7}}
\author{J. R. Houck\altaffilmark{1}}
\author{H. E. Smith\altaffilmark{3}}

\altaffiltext{1}{Department of Astronomy, Cornell University, Ithaca, NY 14853, USA}
\altaffiltext{2}{Spitzer Science Center, 1200 East California Boulevard, Pasadena, CA 91125, USA}
\altaffiltext{3}{Center for Astrophysics and Space Sciences 0424, University of California, San Diego, CA 92093, USA}
\altaffiltext{4}{Department of Astronomy, University of Virginia, P.O. Box 400325, Charlottesville, VA 22904, USA}
\altaffiltext{5}{Astrophysics Group, The Blackett Laboratory, Imperial College, London, UK}
\altaffiltext{6}{Institut d'Astrophysique de Paris, CNRS \& Universit\'e Pierre et Marie Curie, 98bis, bd. Arago, 75014 Paris, France}
\altaffiltext{7}{Astronomy Centre, Department of Physics \& Astronomy, University of Sussex, UK}

\begin{abstract}
We present mid-infrared spectra of thirty two high redshift ultraluminous infrared galaxies, selected via the stellar photospheric feature at rest-frame 1.6$\mu$m, and an observed-frame 24$\mu$m flux of $>$500$\mu$Jy. Nearly all the sample reside in a redshift range of $\langle z \rangle=1.71\pm0.15$, and have rest-frame 1-1000$\mu$m luminosities of 10$^{12.9}$ - 10$^{13.8}$L$_{\odot}$. Most of the spectra exhibit prominent polycyclic aromatic hydrocarbon emission features, and weak silicate absorption, consistent with a starburst origin for the IR emission. Our selection method appears to be a straightforward and efficient way of finding distant, IR-luminous, star-forming galaxies in narrow redshift ranges. There is however evidence that the mid-IR spectra of our sample differ systematically from those of local ULIRGs; our sample have comparable PAH equivalent widths but weaker apparent silicate absorption, and (possibly) enhanced PAH 6.2$\mu$m/7.7$\mu$m and 6.2$\mu$m/11.2$\mu$m flux ratios. Furthermore, the composite mid-IR spectrum of our sample is almost identical to that of local starbursts with IR luminosities of $10^{10}-10^{11}$L$_{\odot}$ rather than that of local ULIRGs. These differences are consistent with a reduced dust column, which can plausibly be obtained via some combination of (1) star formation that is extended over spatial scales of 1-4Kpc, and (2) star formation in unusually gas-rich regions.  
\end{abstract}


\section{Introduction}
First discovered in the 1970s \citep{rie72}, Ultraluminous Infrared Galaxies (ULIRGs, those objects with 1-1000$\mu$m luminosities in excess of $10^{12}$L$_{\odot}$) are among the brightest objects in the low redshift Universe, with luminosities rivalling the bolometric output of quasars. Their immense luminosities, implying astrophysical processes operating on extreme scales, have made ULIRGs the object of intense study over the last thirty years, and in the last decade or so a nascent consensus has developed on their nature. Local ULIRGs are probably `composite' objects; most are powered mainly by a starburst, but a significant fraction ($\sim45\%$) also contain an IR-luminous AGN. This is suggested from several lines of evidence, including optical/UV spectroscopy \citep{vei99,far05}, mid-infrared spectroscopy \citep{gen98,lut98,rig99,arm07,far07b,ima07}, and X-ray observations \citep{fra03,pta03}. Local ULIRGs are also associated almost exclusively with galaxy mergers \citep{far01,bus02,vei02}, and may be involved to some degree in triggering QSOs \citep{san88,tac02,kaw06,zau07}. Reviews of the properties of ULIRGs can be found in \citet{san96}, and more recently in \citet{lfs06}.

Initially, studies of ULIRGs were motivated mostly by their luminosities, as they are rare at low redshifts (with less than 50 examples known over the whole sky at $z<0.2$), and so were not thought to play a fundamental role in the broader picture of galaxy evolution. This perception changed abruptly however, when it was realized that ULIRGs are vastly more numerous at high redshift. First hinted at by spectroscopic followup of {\it Infrared Astronomical Satellite} (IRAS) surveys \citep{hac87,lon90,sau90}, which showed strong evolution in the ULIRG luminosity function with redshift, this was confirmed by surveys with the {\it Infrared Space Observatory} (ISO \citealt{rr97,dol,ver05}) and in the sub-mm \citep{hug,eal,scot,bor,mor}, which showed that there were several hundred ULIRGs {\it per square degree} at $z\gtrsim1$. Later surveys with the {\it Spitzer} space telescope \citep{wer04} showed that ULIRGs contributed an increasing fraction of the comoving bolometric energy density with increasing redshift, reaching a level comparable to the contribution from lower luminosity systems at $z>1$ (\citealt{lef05}, their figure 14). Distant ULIRGs seem to be similar to their low redshift counterparts, in that they appear to be powered by both starburst and AGN activity \citep{far02b,sma03,sma04,ale05,tak06,val07}, and are probably mostly mergers \citep{far02a,cha03,bri07}.

The much higher space density of ULIRGs in the high redshift Universe compared to locally implies that they play an increasingly important role in galaxy evolution with increasing redshift; for example, their high star formation rates make distant ULIRGs excellent candidates for being the rapid growth phases of massive ($\gtrsim2L^{*}$) elliptical galaxies \citep{scot,roc,swi06}. The sheer number of ULIRGs at high redshift however also raises some serious problems, notably for theories for the formation of large-scale structures. These theories usually invoke `biased' hierarchical buildup, in which overdense halos in the dark matter distribution undergo successive mergers to build haloes of increasing mass, with galaxies forming from the baryonic matter in these haloes \citep{col,gran,hatt}. In this scenario, ULIRGs, being the likely antecedents of massive systems, should be found in overdense peaks in the underlying DM distribution, and observations seem to bear this out \citep{bla04,far06a,far06b}. Despite this, the prevalence of ULIRGs at high redshift, and their very high inferred star formation rates, cause major problems for models as they find it difficult to concentrate enough baryons in sufficiently small volumes, and a diverse variety of solutions have been proposed to overcome this \citep{vank,gran2,bau,bow}.

There is therefore a pressing need for both an efficient census of ULIRGs at $z>1$, and a thorough understanding of how these distant ULIRGs form stars. {\it Spitzer} is ideal for undertaking such studies; the Infrared Array Camera (IRAC, \citealt{faz04}) and Multiband Imaging Photometer for Spitzer (MIPS, \citealt{rie04}) imaging instruments, and the Infrared Spectrograph (IRS, \citealt{hou04}) all offer dramatic improvements in sensitivity and resolution compared to previous generation facilities. In this paper, we use the IRS to study 32 ULIRGs at $z\gtrsim1$, selected using IRAC and MIPS photometry. We quantify the efficiency of the selection method and use mid-IR spectral features to set constraints on the nature of their star formation. An overview of the sample selection and a description of the observations is given in \S\ref{sectobs}. Results are summarized in \S\ref{takeaguess} with discussion and conclusions given in \S\ref{discuss} and \S\ref{conc} respectively. We assume a spatially flat cosmology with $H_{0}=70$ km s$^{-1}$ Mpc$^{-1}$, $\Omega=1$, and $\Omega_{m}=0.3$. Unless otherwise stated, the term `Infrared (IR) luminosity' refers to the total luminosity integrated over 1-1000$\mu$m in the rest-frame.

\section{Analysis}\label{sectobs}

\subsection{Sample Selection}\label{select}
The basis of our selection is an apparent emission feature at rest-frame 1.6$\mu$m, produced by thermal emission from late-type stars and enhanced by an apparent emission feature caused by H$^{-}$ ions in the atmospheres of giant stars\footnote{H$^{-}$ is a major source of opacity in the atmospheres of such stars, but shows a minimum in opacity at around 1.6$\mu$m \citep{joh88}, leading to a broad apparent emission feature at this wavelength}. This 1.6$\mu$m `bump' is almost ubiquitous in the integrated spectra of galaxies. Previous authors \citep{sim99,saw02} have investigated the use of this feature in deriving photometric redshifts for galaxies using the Spitzer IRAC channels, and conclude that it is likely to be effective. A 1.6$\mu$m bump also implies that any AGN continuum is not significant in the rest-frame near-IR, although it does not rule out the possibility of a low luminosity and/or very obscured AGN. Therefore, combining the presence of the 1.6$\mu$m feature with a prior on IR luminosity should select high redshift ULIRGs that harbor active star formation. A complete description of the selection method can be found in Lonsdale et al (2008, in preparation). Angular clustering measurements for large samples of `bump' selected ULIRGs can be found in \citet{far06a} (see also \citealt{oli04,wad07,tor07}), and millimeter-wave fluxes and analysis can be found in \citet{lon07b}. Other studies that include small numbers of `bump' sources include IRS spectroscopy \citep{wee06}, and optical spectroscopy \citep{ber07}. Other studies of the IRAC colors of SWIRE galaxies can be found in \citet{dava,davb}.

We select our sources from the SWIRE \citep{lon03} Lockman field, which spans 10.2 square degrees, and has been imaged in all seven IRAC and MIPS bands to 5$\sigma$ depths of 4.2$\mu$Jy, 7.5$\mu$Jy, 46.1$\mu$Jy, 47.3$\mu$Jy, 210$\mu$Jy, 18mJy and 108mJy respectively. We first excluded all sources brighter than $m_{r}=23$ to remove local and low redshift galaxies, and all sources fainter than 500$\mu$Jy at 24$\mu$m to ensure that spectra could be obtained with the IRS in reasonable integration times\footnote{The implied 24/optical color-cut is substantially more lenient than that of \citealt{hou05} and \citealt{yan07}, which would have excluded an Arp 220 type galaxy at z=2.}. We then selected all sources that exhibited excess emission in the 4.5$\mu$m channel, i.e. where $f_{3.6}<f_{4.5}>f_{5.8}$ and $f_{4.5}>f_{8.0}$ (or `bump 2' sources). This resulted in a total of 350 sources. We then randomly selected 31 objects for observation with IRS. We also included one further object in the SWIRE ELAIS N1 field that satisfies all of our selection criteria, and was observed as part of a complementary program. 

The complete sample of 32 objects is listed in Table \ref{sample}. The IRAC and MIPS fluxes in this table are from the latest SWIRE data release (the `SWIRE3' catalogs). The sources were however originally selected using the previous generation, `SWIRE2' catalog. The changes to the IRAC pipeline between these two catalog versions resulted in flux differences between the catalogs of only a few percent, but our original selection did not demand that the bump was detected with any statistical robustness, and in some cases the band 2 excess was not significant within the errors. As a result, some of our sample no longer display an excess in the 4.5$\mu$m channel using the SWIRE3 catalogs. The IRAC fluxes of all of the sample are however consistent within the errors with an excess in the 4.5$\mu$m channel, and for simplicity we treat all of the sources in Table \ref{sample} as `bump 2' sources for the remainder of this paper.

\subsection{Observations}
Each source was observed with the Long-Low module onboard the IRS ($10.5\arcsec\times168\arcsec$, 5.1\arcsec\ pix$^{-1}$, R$\sim57-126$) in both first (LL1, 19.5$\mu$m - 38.0$\mu$m) and second (LL2, 14.0$\mu$m - 21.3$\mu$m) order, as part of Spitzer program 30364 and (for source 32) program 15. The targets were placed in the center of each slit using the blue peak-up array. Each target was observed for eight ramp cycles, with an individual ramp time of 120 seconds\footnote{Details for each observation can be found by referencing the AOR keys given in Table \ref{sample} within the {\it Leopard} software, available from the Spitzer Science Center}.

The data were processed through the {\it Spitzer} Science Center's pipeline reduction software (version 15.3), which performs standard reduction tasks such as ramp fitting and dark current subtraction, and produces Basic Calibrated Data (BCD) frames. Starting with these frames, we flagged rogue and otherwise `bad' pixels using the {\em irsclean}\footnote{This tool is available from the SSC website: http://ssc.spitzer.caltech.edu} tool, which uses a pixel mask for each campaign to first flag and then replace rogue pixels. The individual frames at each nod position were then combined into a single image using the SMART software package \citep{hig}. Sky background was removed from each image by subtracting the image for the same object taken with the adjacent order (i.e. `order-order' sky subtraction). One-dimensional spectra were then extracted from the images using the SPICE software package using `optimal' extraction and default parameters for the Long-Low modules. This results in separate spectra for each nod and for each order, all of which must be combined to give the final spectrum. The spectra for each nod were inspected; features present in only one nod were treated as artifacts and removed manually. The two nod positions were then combined. In most cases the two orders differed in flux by $\sim20\%$; we first scaled LL2 to LL1, and then scaled the resulting spectrum to the 24$\mu$m flux. Finally, the first and last 4 pixels on the edge of each order, corresponding to regions of decreased sensitivity on the array, were removed to give the final spectrum for each object.

\section{Results}\label{takeaguess}
\begin{figure}
\begin{minipage}{180mm}
\includegraphics[angle=90,width=170mm]{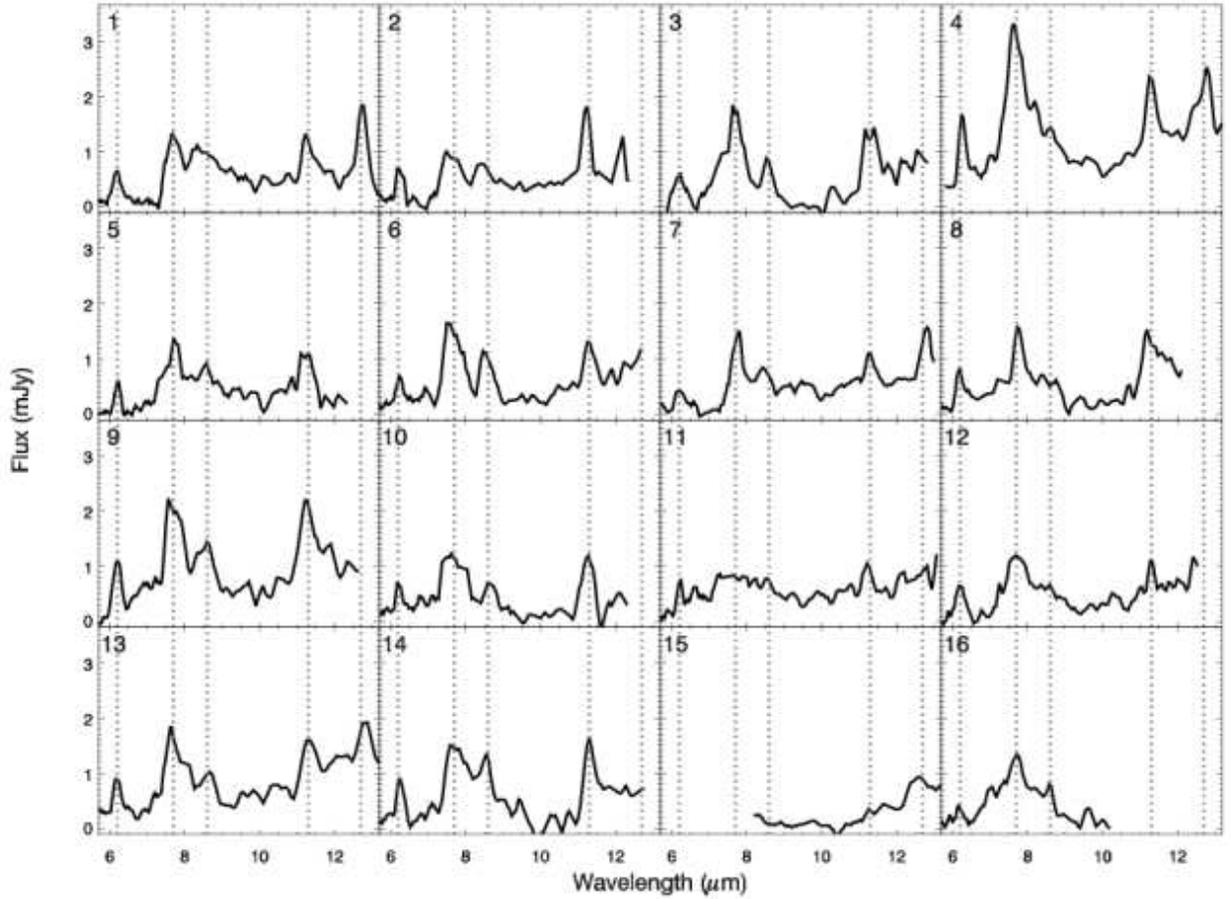}
\end{minipage}
\caption{Individual spectra, plotted in the rest-frame using the spectroscopic redshifts in Table \ref{measurements}. The number in the top left of each panel corresponds to the ID number in the tables. The vertical dotted lines mark the wavelengths of the 6.2$\mu$m, 7.7$\mu$m, 8.8$\mu$m, 11.2$\mu$m and 12.7$\mu$m PAH features.
\label{spectraa}}
\end{figure}

\begin{figure}
\begin{minipage}{180mm}
\includegraphics[angle=90,width=170mm]{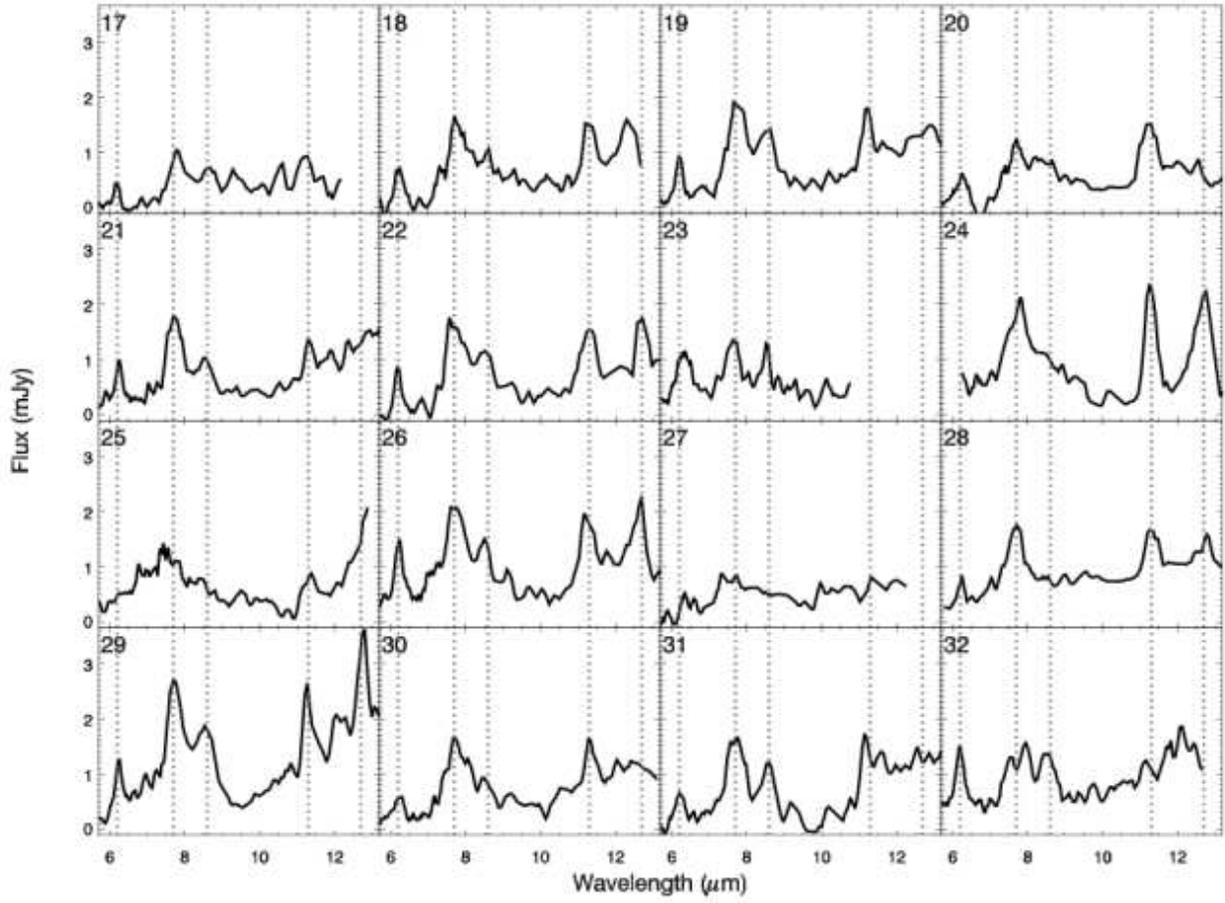}
\end{minipage}
\caption{Individual spectra for our sample, continued from Figure \ref{spectraa}. 
\label{spectrab}}
\end{figure}

The spectra are presented in Figures \ref{spectraa} and \ref{spectrab}. Redshift data are presented in Table \ref{measurements}, and measurements of spectral features are given in Table \ref{pahsils}.

\begin{figure}
\begin{minipage}{180mm}
\includegraphics[angle=90,width=85mm]{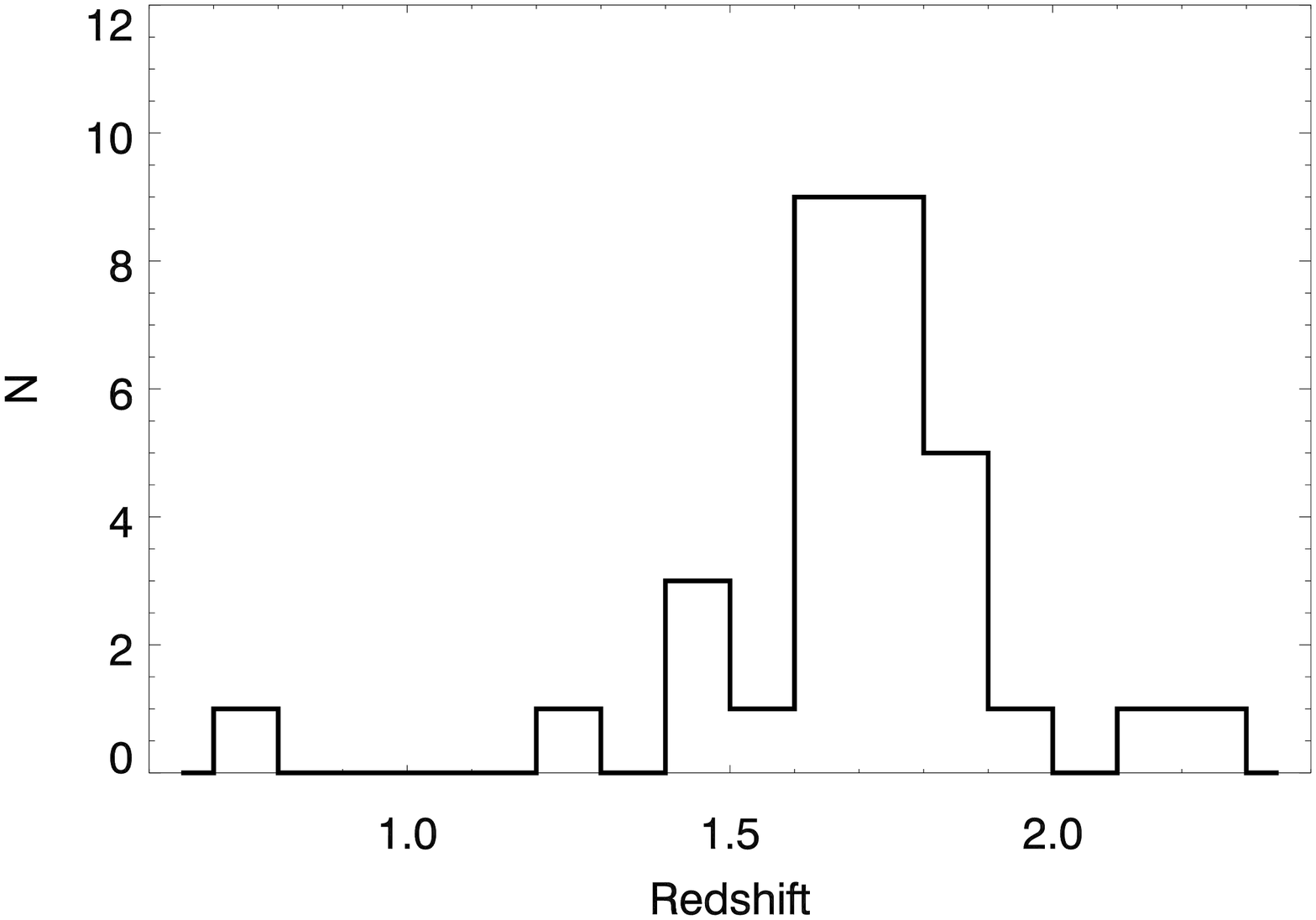}
\includegraphics[angle=90,width=85mm]{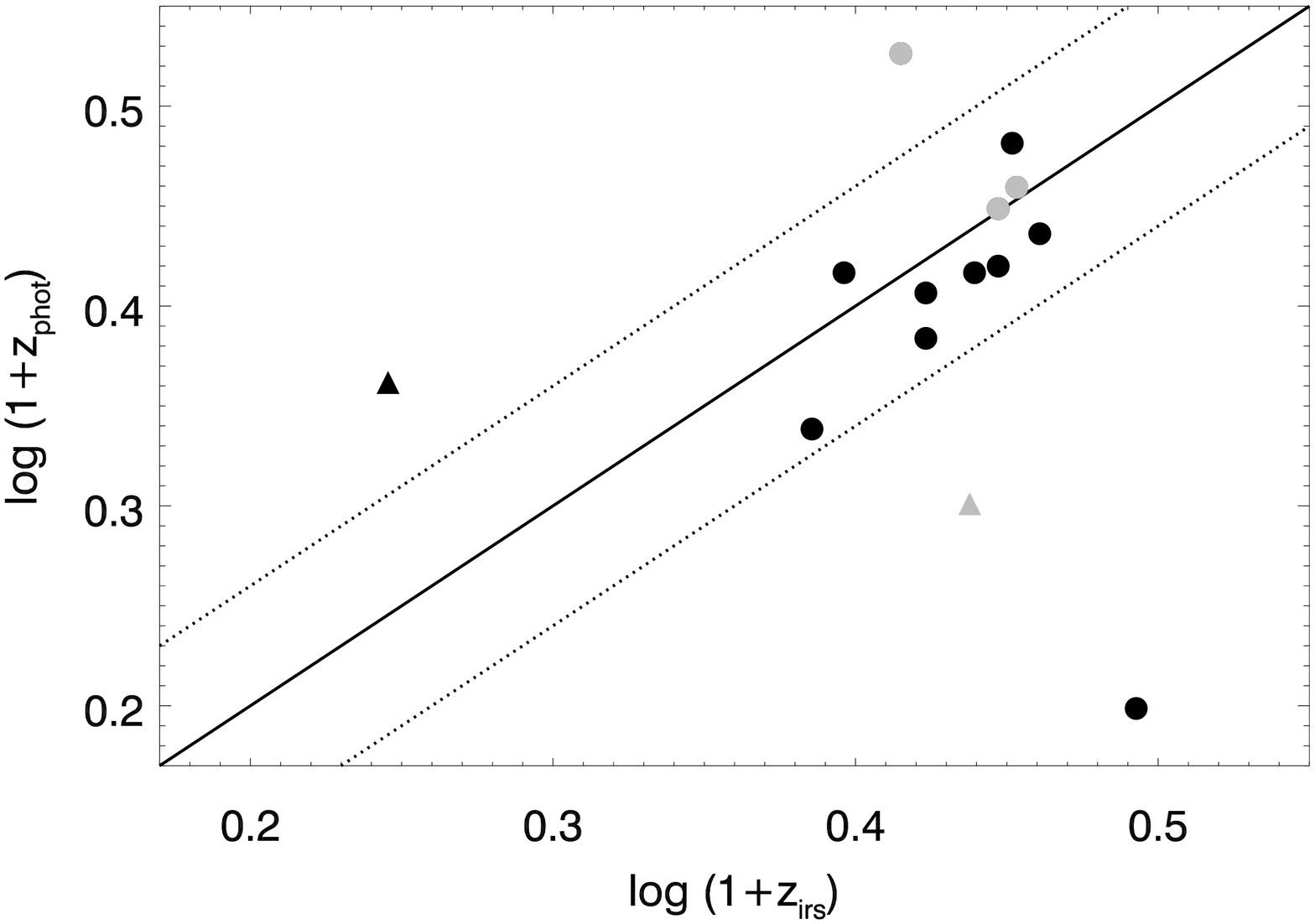}
\end{minipage}
\caption{{\it Left panel:} Redshift histogram, using the redshifts derived from the IRS spectra. Nearly all of the sample is confined to a narrow redshift range of $1.4<z<1.9$. {\it Right panel:} Comparison between the IRS redshifts, and the photometric redshifts from \citet{rr07}, listed in Table \ref{measurements}. The dotted lines denote a deviation of 0.06 in log(1+z), the boundary defined as a `catastrophic failure' by \citet{rr07}. Sources plotted as triangles have an uncertain IRS redshift. Sources plotted in grey have a photometric redshift derived from three bands. 
\label{zphotzspec}}
\end{figure}

We derive spectroscopic redshifts from broad emission features at 6.2$\mu$m, 7.7$\mu$m, 8.6$\mu$m, 11.2$\mu$m and 12.7$\mu$m, attributed to bending and stretching modes in neutral and ionized PAH molecules (the 12.7$\mu$m feature likely also contains a contribution from the [NeII]$\lambda$12.814 fine-structure line), and/or a broad absorption feature at 9.7$\mu$m arising from large silicate dust grains. In most cases (29/32) an unambiguous redshift can be determined from the PAH features; the associated uncertainty on these redshifts is governed by the variations in PAH peak wavelengths seen in local galaxies, and is of order $\Delta z=0.02$. In three cases however (sources 11, 15 and 27) the spectral features cannot be reliably identified. For these three sources we derive tentative redshifts based on what are plausibly PAH or silicate features. In these cases the error on the redshift is larger, $\Delta z\simeq 0.3$, and the redshift itself should be treated with caution. A histogram of the redshifts is plotted in the left panel of Figure \ref{zphotzspec}. All of the sample lie in the redshift range $0.8<z<2.3$, with all but three lying within $1.4<z<1.9$. The redshift distribution is well fitted by a Gaussian centered at $z$=1.71 and with a Full Width at Half Maximum of $\Delta$z$\simeq$0.35.

An accurate measure of the IR luminosities is difficult due to the paucity of data at far-IR wavelengths. Therefore, we initially simply estimate the range in IR luminosity within which our sample likely resides, and defer estimates of IR luminosities based on the IRS spectra to \S\ref{nature}. To do this, we normalize a suite of radiative transfer models for starbursts \citep{efs00} and AGN \citep{efs95} to the 24$\mu$m fluxes, and reject those models that exceed the IRAC fluxes, or 70$\mu$m or 160$\mu$m upper limits. The IR luminosities derived from the remaining models then constitute the probable range in IR luminosity of our sample. The resulting IR luminosity range is $10^{12.6} - 10^{13.6}$L$_{\odot}$, placing our sample at the upper end of the IR luminosity range seen for low-redshift ULIRGs

The sample is faint at 24$\mu$m, hence our spectra are of low signal-to-noise, and detailed spectral diagnostics are not possible. Most of the spectra however show one or more PAH features at 6.2$\mu$m  7.7$\mu$m  8.6$\mu$m, 11.2$\mu$m and (depending on redshift) 12.7$\mu$m. The PAH fluxes and equivalent widths (EWs) were computed by integrating the flux above a spline interpolated local continuum fit (for a description of the method see \citealt{bra06} and \citealt{spo07}). Due to the low S/N and restricted wavelength range of the spectra, we cannot correct for water ice and/or aliphatic hydrocarbon absorption, however the effect of this lack of correction is likely to be insignificant. The errors on the EWs are large because the method used to measure them requires a spline  fit to the underlying continuum, and the continua of our sample are in most cases only weakly detected. This method does however offer the advantage that our EWs can be compared directly to those of local IR-luminous galaxies as measured by Spitzer \citep{spo04,wee05,bra06,arm07,spo07}.

We derive star formation rates from the combined luminosity of the 6.2$\mu$m and 11.2$\mu$m PAH features, via the formula from \citet{far07b}:

\begin{equation}
\stackrel{.}{M_{\odot}} = 1.18\times10^{-48}L_{P}\label{pahstarconc}
\end{equation}

\noindent where $\stackrel{.}{M_{\odot}}$ is the star formation rate in solar masses per year, and $L_{P}$ is the PAH 6.2$\mu$m + 11.2$\mu$m luminosity in Watts. The resulting star formation rates are listed in Table \ref{pahsils}. As described in \citet{far07b}, the errors on these star formation rates are of order $50\%$.

We measured the strengths of the silicate features via:

\begin{equation}
S_{sil} = ln\left(  \frac{F_{obs}(9.7\mu m)}{F_{cont}(9.7\mu m)} \right)\label{silstrength}
\end{equation}

\noindent where $F_{obs}$ is the observed flux density at rest-frame 9.7$\mu$m, and $F_{cont}$ is the underlying continuum flux at rest-frame 9.7$\mu$m deduced from a spline fit to the continuum on either side. A complete description of the method used to measure $S_{sil}$ can be found in \citet{spo07,lev07} and \citet{sir07}. The silicate feature is not detected (at $>3\sigma$) in any object individually. We also see no evidence for molecular Hydrogen lines or any fine-structure emission lines in the individual spectra. 

\begin{figure}
\begin{minipage}{180mm}
\includegraphics[angle=90,width=170mm]{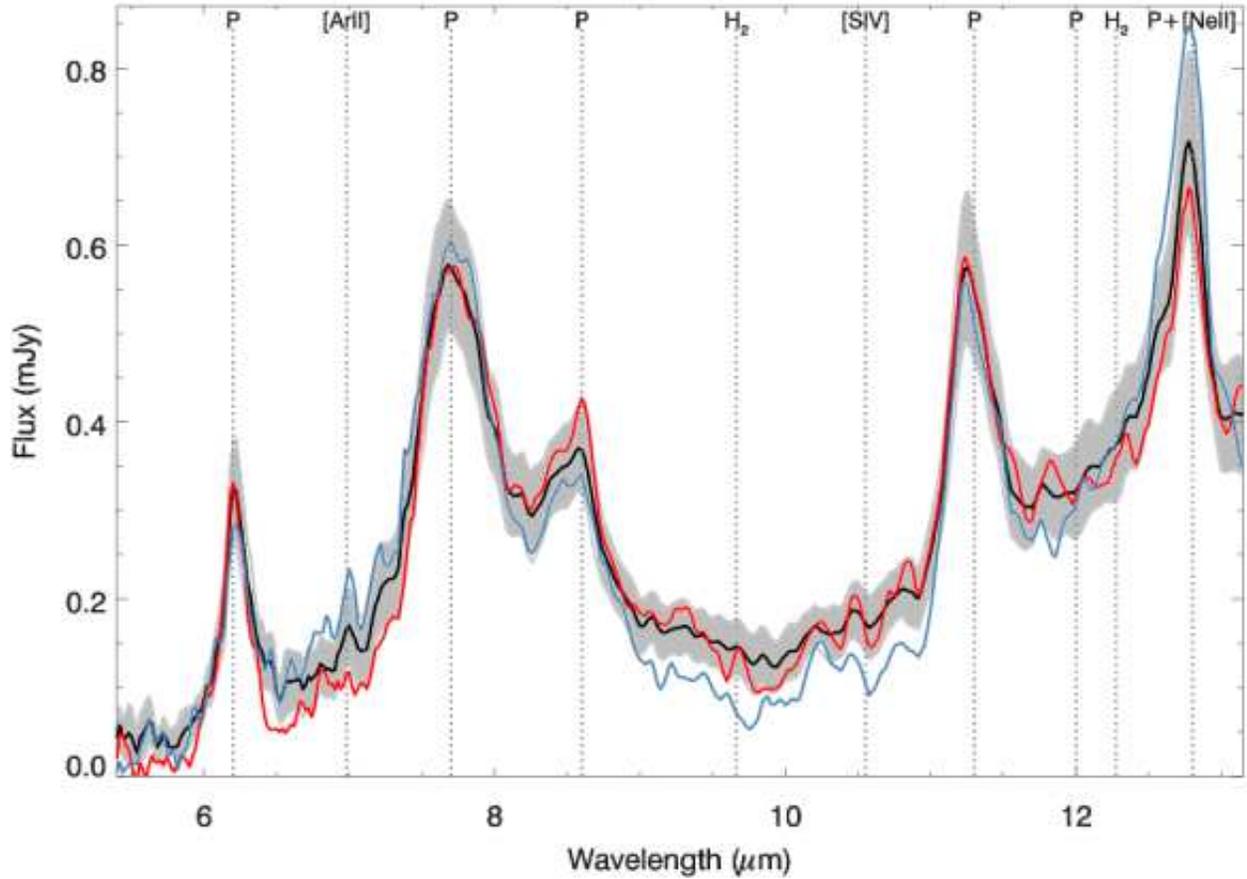}
\end{minipage}
\caption{Averaged spectra. The average of all 32 objects is plotted in black, with the grey shaded region indicating the $1\sigma$ dispersion. The average of the ten objects with the largest PAH 6.2$\mu$m EWs is plotted in red, and the average of the ten objects with the largest silicate strengths is plotted in blue. A `P' denotes the wavelength of a PAH feature.
\label{composite_all}}
\end{figure}

The average spectrum for all 32 objects is shown in Figure \ref{composite_all}. All of the PAH features between 6.2$\mu$m and 12.7$\mu$m are prominent. There is also evidence for two fine-structure lines; the 12.7$\mu$m PAH feature appears asymmetric, likely due to the presence of [NeII]$\lambda$12.814, and there is a weak detection of what may be [ArII]$\lambda$6.99. Both these lines are commonly seen in local ULIRGs \citep{gen98,far07b}. Conversely, the average spectrum shows no evidence for significant apparent silicate absorption. We also see no signs of the H$_{2}$,S(3)$\lambda$9.662 and H$_{2}$,S(2)$\lambda$12.275 lines, which are both common, though weak, in local ULIRGs, or the [SIV]$\lambda$10.511 line. These results do not change appreciably if we selectively coadd the spectra, examples of which are also shown in Figure \ref{composite_all}. The average of the ten spectra with the largest PAH 6.2$\mu$m EWs shows no apparent silicate absorption, no molecular Hydrogen features, but strong PAH features with an asymmetric 12.7$\mu$m PAH feature. The average of the ten spectra with the `strongest' silicate absorption features is similar; there is a hint of absorption, but it is not detected with any statistical significance.

\section{Discussion}\label{discuss}

\subsection{Redshifts and luminosities}
The redshifts of our sample have a remarkably tight distribution. All of the sample lie in the range $0.8<z<2.3$, with all but three lying within $1.4<z<1.9$. If we exclude these three outliers, then the mean redshift for our sample is $\langle z \rangle=1.71\pm0.15$ (assuming FWHM$=(2\sqrt{2ln2})\sigma$). This is significantly tighter than the redshift distribution of sub-mm selected galaxies, which have $\langle z \rangle = 2.2 \pm 0.5$ \citep{cha05}, and the redshift distribution for Lyman break galaxies, which have $\langle z \rangle=2.96\pm0.29$ \citep{stei03}, though a larger sample would be needed to confrm this. Our redshift distribution is much tighter than that of sources found by selecting on 24$\mu$m/0.7$\mu$m flux ratio \citep{yan07}, which span $0.5<z<3.0$. Combined with the constraint that the IR luminosities of our sample must (for any reasonable SED shape) be in excess of 10$^{12.5}$L$_{\odot}$, we conclude that our selection method is an excellent way of finding distant ULIRGs in specific, narrow redshift intervals. This narrow redshift range likely arises due to a combination of two factors; the demand of an excess flux at observed frame 4.5$\mu$m favors sources with a rest-frame 1.6$\mu$m opacity feature at $1.4<z<2.0$, and the demand that sources be bright at observed-frame 24$\mu$m biases towards sources whose redshifts place the 7.7$\mu$m and 8.6$\mu$m PAH features within the MIPS 24$\mu$m band. This method offers many advantages over other methods for finding high redshift ULIRGs. For example, it avoids all of the observational difficulties in assigning optical counterparts to sub-mm selected ULIRGs (see for example \citet{dun04}). Remarkably, we have arrived at this redshift distribution without demanding that the `bump' be detected with any statistical robustness, and it is notable that none of the seven sources that `lost' their bump in the 4.5$\mu$m channel due to reprocessing (Table \ref{measurements}) are unusual in redshift or spectral shape. 

Given the difficulties in obtaining accurate redshifts for distant ULIRGs \citep{sim04,cha05,ber07,are07}, we examine the accuracy of photometric redshifts for our sample obtained by \citet{rr07} and listed in Table \ref{measurements}. Given the high redshifts and limited optical photometry, the photometric redshifts are surprisingly good. `Reliable' solutions (those with $\chi^{2}_{red}<10$ and detected in at least three of the $Ugriz$ bands) are obtained for 14/32 objects; these are plotted in the right panel of Figure \ref{zphotzspec}. Of those 14, four are classified as `catastrophic' failures, i.e. those where the photometric redshift deviates from the spectroscopic redshift by more than $6\%$ in log(1+z) (although two of these four are sources 11 and 15, which have uncertain spectroscopic redshifts), while ten are classified as successes. We conclude that the \citet{rr07} photometric redshift method remains effective even for faint, high redshift sources. Finally, we note that, of the seven objects in Table \ref{measurements} with ambiguous bump classifications, six have photometric and spectroscopic redshifts that agree very well. It seems therefore that, for the sources with an ambiguous bump classification, most are indeed `bump 2' sources.

\subsection{What powers the infrared emission?}
The IRS spectra provide the best diagnostics of the power source behind the IR emission. The presence of prominent PAH features in 29/32 objects is strong evidence that $\gtrsim 90\%$ of our sample harbor powerful starbursts. The star formation rates computed using Equation \ref{pahstarconc} and listed in Table \ref{pahsils} are all extremely high. If we extrapolate these star formation rates to IR luminosities via the relation given by \citet{ken98}\footnote{\citet{ken98} derive a relation between star formation rate and rest-frame 8-1000$\mu$m luminosity rather than 1-1000$\mu$m luminosity, but the difference in wavelength range is unlikely to be significant}, then these luminosities lie in the same range as those determined in \S\ref{takeaguess}. We infer that dust heated by star formation can account for all of the IR emission from the sources that show strong PAHs, and therefore that most of our sample are consistent with being pure starbursts. Interestingly, the strong PAHs and high degree of homogeneity of our sample make them similar to sub-mm selected ULIRGs \citep{men07,pop07}, suggesting that we may be drawing from the same parent population.

We cannot, however, rule out the presence of an AGN in any of our sample. It is unlikely that an AGN contributes significantly in the rest-frame near-IR, as we see no power law continua in the IRAC bands (see e.g. \citealt{alo06,pol06}), but it is possible that some of our sample contain mid-IR luminous AGN, even in those objects with prominent PAHs. An example of such an object at low redshift is NGC 6240, which shows strong PAHs accompanied by an AGN continuum and the [NeV]$\lambda$14.322 fine-structure line\footnote{[NeV]$\lambda$14.32  has $E_{ion}=97.1$ eV, and models indicate that it is unlikely to be strong in galaxies without an AGN \citep{voi92}. Observationally, [NeV]$\lambda$14.32 is occasionally seen in planetary nebulae and young supernova remnants \citep{oli99}, but is not seen in the integrated spectra of starburst galaxies. It is however strong in the spectra of AGN \citep{lut98}} \citep{arm06}. Furthermore, previous authors have noted correlations between shallow ($S_{sil}<0.8$) or deep ($S_{sil}>2.4$) silicate features, and the presence of AGN; \citet{far07b} show that ULIRGs with shallow silicate features usually also show the [NeV]$\lambda$14.322 line, \citet{hao07} find that QSOs and Seyfert 1 galaxies often show the silicate feature in emission, and \citet{saj07} find that many high redshift ULIRGs with deep silicate absorption features also show powerful radio jets. The shallow silicate features of our sample are therefore consistent with (though not supportive of) the presence of an AGN.

\begin{figure}
\begin{minipage}{180mm}
\includegraphics[angle=90,width=170mm]{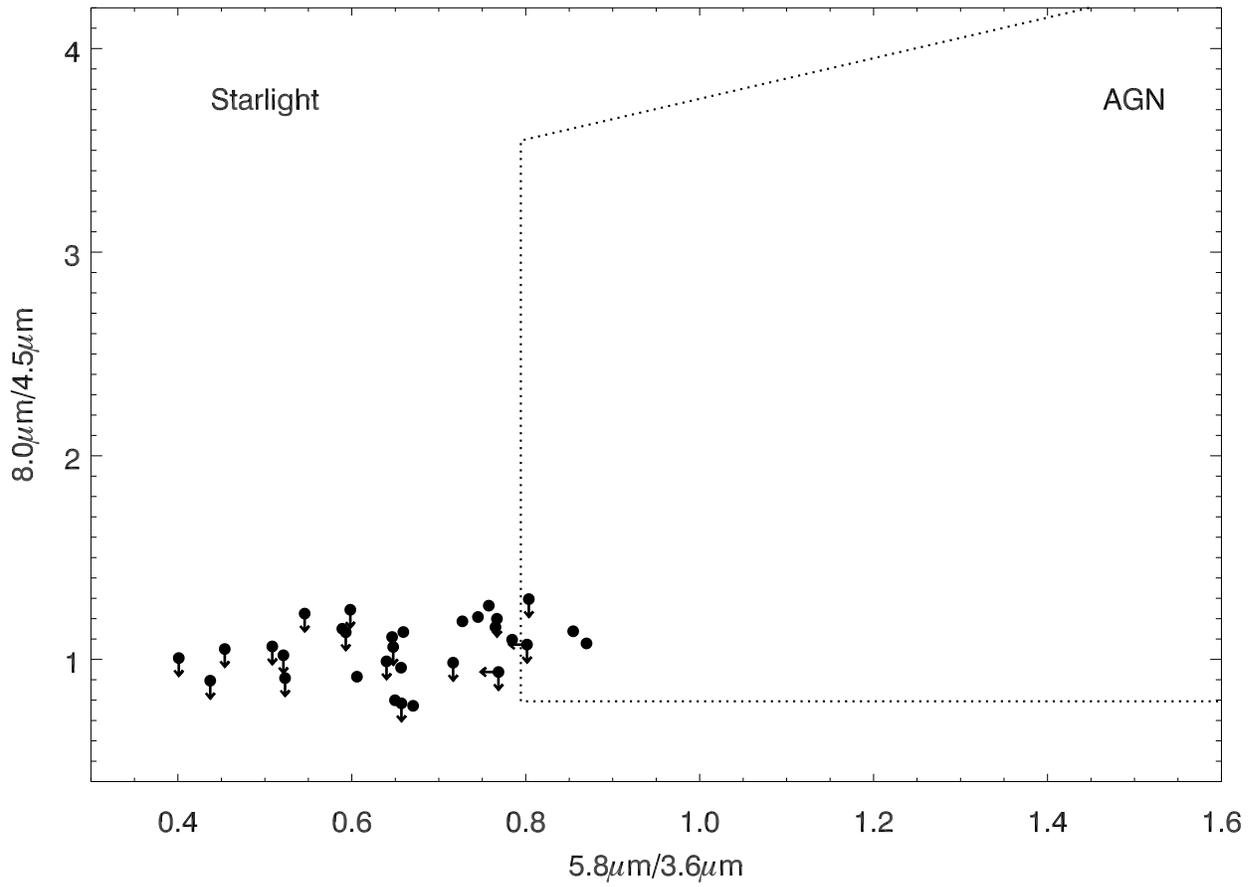}
\end{minipage}
\caption{IRAC color-color plot. The `wedge' is the selection region for sources that exhibit AGN-like IRAC colors, defined empirically using IRAC observations of QSOs and Seyfert galaxies \citep{lac04}. 
\label{colcol}}
\end{figure}

Finally, we examine the IRAC colors as a diagnostic of the IR emission. In Figure \ref{colcol} we plot 5.8$\mu$m/3.6$\mu$m color against 8.0$\mu$m/4.5$\mu$m color, together with the selection `wedge' for AGN from \citet{lac04}, defined empirically using the Spitzer colors of SDSS QSOs. This diagnostic is not, on its own, particularly useful, as our 4.5$\mu$m `bump' selection already excludes much of the AGN selection region, but it serves as a useful check. Nearly all of our sample lie outside the AGN selection region, and are consistent with the positions of the bulk of the galaxies in the main Spitzer First Look Survey (FLS) dataset. Our sample lies some distance from the SDSS QSOs and Sy1 galaxies in figure 1 of \citet{lac04}. Based solely on the IRAC colors, the rest-frame near-IR emission from our sample is consistent with arising from starlight. 

\subsection{The nature of star formation in distant ULIRGs}\label{nature}

\subsubsection{PAH flux ratios}\label{pahfluxsect}
\begin{figure}
\begin{minipage}{180mm}
\includegraphics[angle=90,width=85mm]{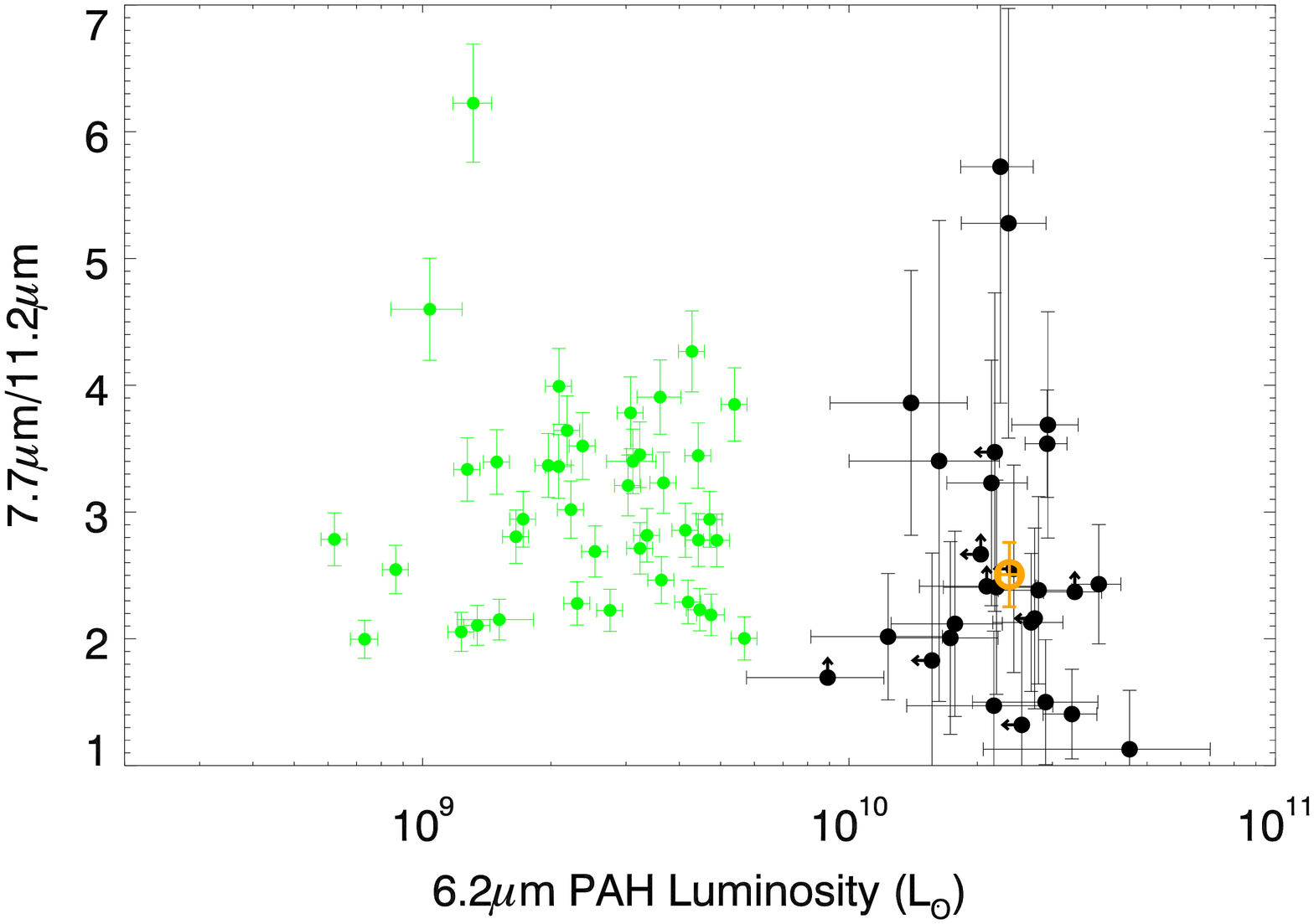}
\includegraphics[angle=90,width=85mm]{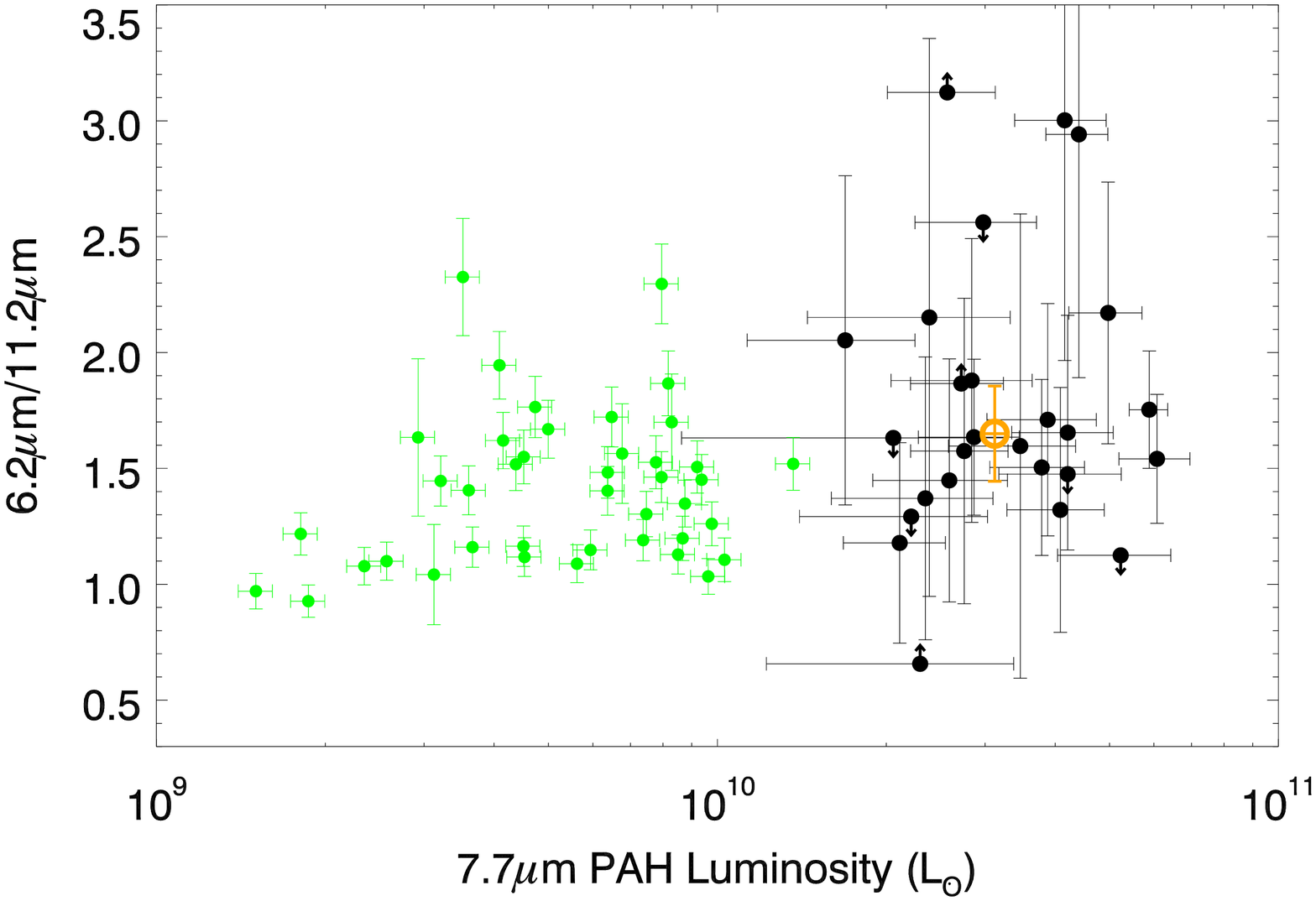}
\end{minipage}
\caption{PAH luminosity vs PAH flux ratio plots. Our sample is plotted in black, and local ULIRGs \citep{spo07} are plotted in green. The large orange point is the mean value for our sample. 
\label{pahlumratio}}
\end{figure}

The prevalence of starbursts in our sample means we can investigate the nature of star formation in high redshift ULIRGs. We start by considering diagnostics based solely on the PAH features. In Figure \ref{pahlumratio} we plot PAH luminosities against PAH flux ratios. The  6.2$\mu$m and 7.7$\mu$m PAH features are extraordinarily luminous, lending further weight to the conclusion that the starbursts in our sample are more intense than seen in any local ($z<0.2$) ULIRG. The dispersion in PAH luminosity is perhaps slightly greater for our sample, but not significantly so given the sizes of the error bars. 

\begin{figure}
\begin{minipage}{180mm}
\includegraphics[angle=90,width=85mm]{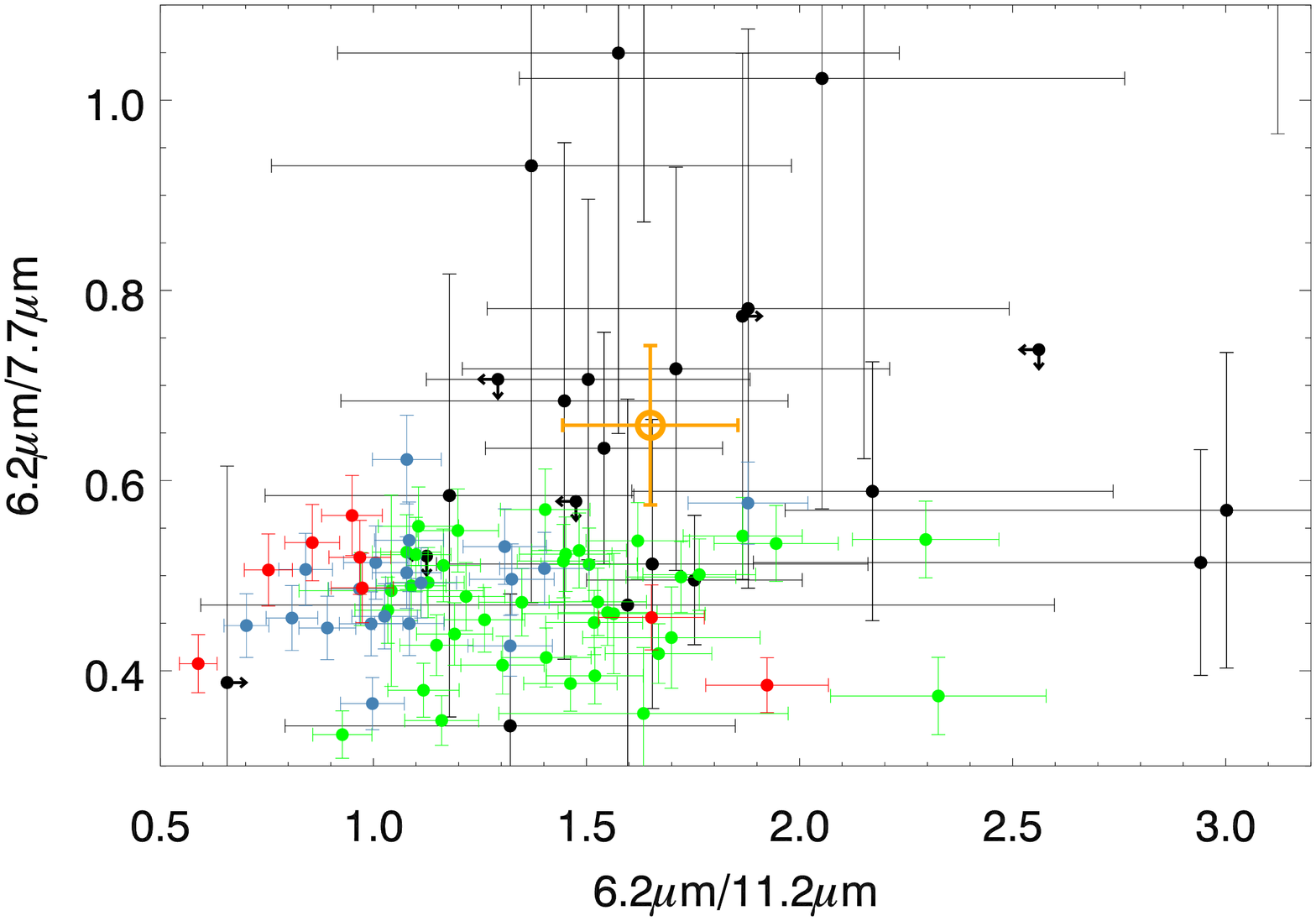}
\includegraphics[angle=90,width=85mm]{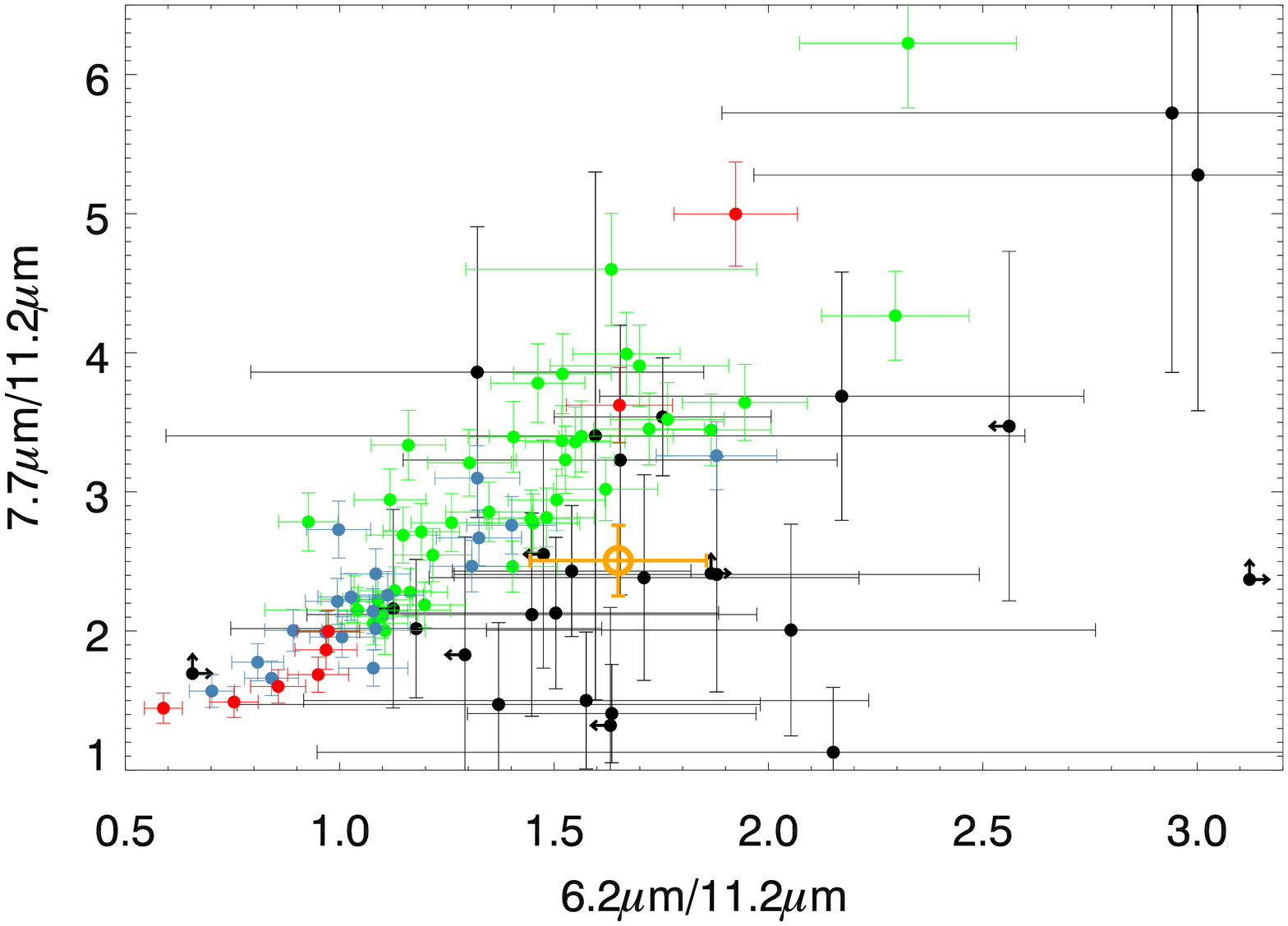}
\end{minipage}
\caption{PAH flux color-color plots. Our sample is plotted in black, local ULIRGs \citep{spo07} are plotted in green, and local `classical' AGN \citep{wee05} and starbursts \citep{bra06} with IR luminosities in the range 10$^{10}$L$_{\odot}$ - 10$^{11.5}$L$_{\odot}$ are plotted in red and blue, respectively. The large orange point is the mean value for our sample.
\label{pahfluxratio}}
\end{figure}

We see a more interesting picture if we consider PAH color-color plots (Figure \ref{pahfluxratio}). Considering first only the low redshift sources;  local IR-luminous galaxies show a distinct `environmental segregation' with IR luminosity, corresponding to elevated 6.2$\mu$m/11.2$\mu$m and 7.7$\mu$m/11.2$\mu$m PAH ratios in ULIRGs compared to lower luminosity starbursts and AGN. Such a segregation is also seen in HII regions and Photo-Dissociation regions, and the reasons for it are not clear \citep{verme02,leb07}. The two most likely reasons are PAH ionization and PAH molecular structure. A higher fraction of ionized PAHs gives rise to stronger PAH features at shorter wavelengths, as does a higher fraction of open, uneven PAH molecules compared to smooth, compact PAH molecules (see figure 9 of \citet{hon01} for diagrams of these PAH features). The drivers behind PAH molecular structure are unknown, but PAH ionization fractions can be affected in two ways; (1) a harder interstellar radiation field (ISRF) would ionize more PAHs\footnote{though \citet{smi07} note that a hard ISRF from a low-luminosity AGN may actually suppress PAHs in the 5-8$\mu$m range}, and (2) decreasing the electron density would decrease the recombination rate, leading to a higher fraction of ionized PAHs. It is therefore likely that local ULIRGs have either a harder ISRF, a lower electron density, or a higher fraction of open, uneven PAH molecules than do lower luminosity systems. 

Turning now to the bump sources; our sample appears to lie preferentially toward one `edge' of the local ULIRG distribution, corresponding to even further elevated 6.2$\mu$m PAH emission compared to the 7.7$\mu$m and 11.2$\mu$m features, though the error bars are large even for the average bump point. These two diagrams alone suggest that the star formation in our sample {\it may} differ in some way from that seen in local ULIRGs, although it is difficult to say what the physical origin of those differences could be. One possibility is a lower extinction level in our sample compared to the mean for low redshift ULIRGs.

\subsubsection{PAH fluxes \& Silicate strengths}\label{pahfluxsilicates}
\begin{figure}
\begin{minipage}{180mm}
\includegraphics[angle=90,width=170mm]{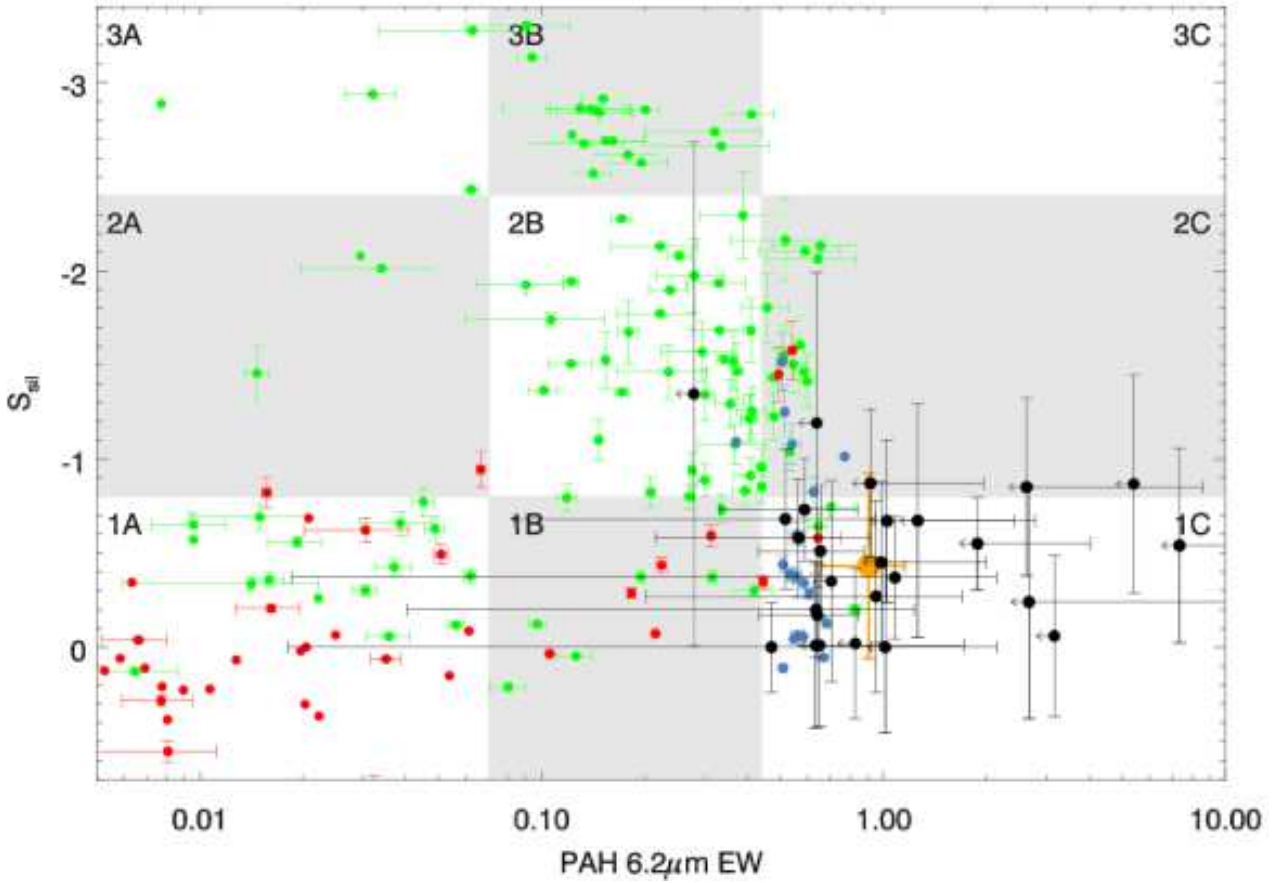}
\end{minipage}
\caption{The `Fork' diagram of \citet{spo07}. Our sources are plotted in black, local ULIRGs are plotted in green, while local starbursts \citep{bra06} and AGN \citep{wee05} with IR luminosities in the range 10$^{10} - 10^{11}$L$_{\odot}$ are plotted in blue and red respectively. The large orange point is the average for all the bump sources. Our sources occupy a small region, mostly the 1C class. This contrasts with local ULIRGs, which are found in nearly all classes. 
\label{fork}}
\end{figure}

\begin{figure}
\begin{minipage}{180mm}
\includegraphics[angle=90,width=170mm]{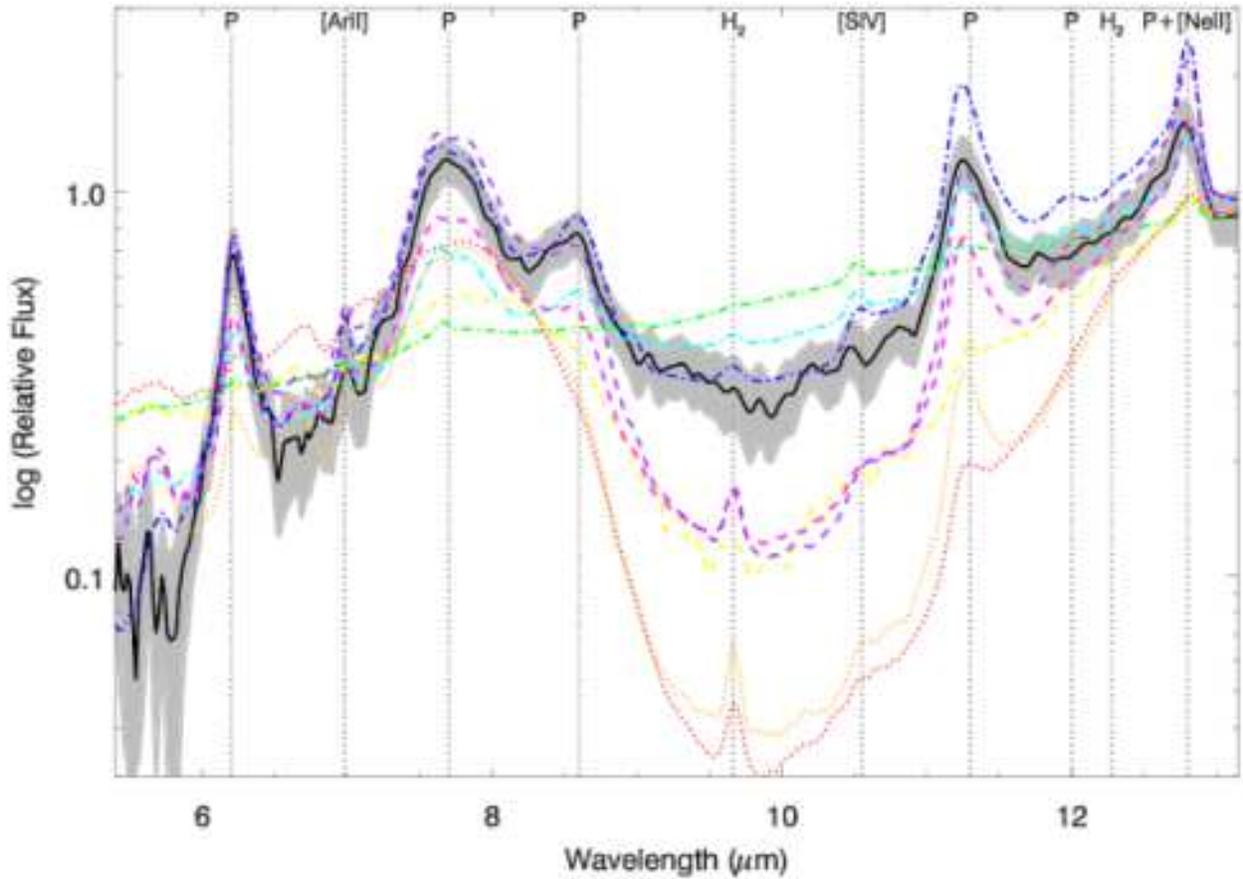}
\end{minipage}
\caption{Comparison between our averaged rest-frame spectrum (black, with 1$\sigma$ errors denoted by the grey region) and averaged spectra for classes in the `Fork' diagram of \citet{spo07}, normalized at 14$\mu$m; class 1A (green), class 1B (cyan), class 1C (blue), class 2A (yellow), class 2B (magenta), class 2C (purple), class 3A(red), and class 3B (orange). \label{comp_fork}}
\end{figure}

We move on to consider diagnostics based on both the PAH emission features and the 9.7$\mu$m silicate absorption feature. In Figure \ref{fork} we plot our sample on the `fork' diagram of \citet{spo07}. Our sources occupy a small region on the lower right of this plot, corresponding almost entirely to class 1C, with a small number in classes 2C and 2B, though the large error bars on the PAH EWs mean our sources are in many cases also consistent with class 1A/1B. These classes label our sources as PAH dominated, with relatively weak apparent silicate absorption. To check whether this is just a chance similarity between these specific spectral features, we plot in Figure \ref{comp_fork} the averaged spectrum from Figure \ref{composite_all} against the averaged spectra for the classes in the fork diagram. Clearly, our averaged spectrum closely resembles the average spectrum for the class 1C sources.

Figure \ref{fork} raises interesting questions on the nature of starburst activity in our sample. As described in \S\ref{select}, our selection method biases towards ULIRGs harboring ongoing star formation, and selects against those ULIRGs containing a rest-frame near-IR luminous AGN. Our sample should therefore resemble the majority of local ULIRGs, as most low redshift ULIRG are powered mainly by star formation (e.g. \citealt{far03}). Furthermore, our demand for an excess flux in one of the IRAC channels should not remove many `local type' ULIRGs from the sample, as many local ULIRGs have a rest-frame SED that either has a near-IR excess, or is not obviously a power law. Moreover, previous studies of high redshift, L$_{IR}\simeq10^{13}$L$_{\odot}$ ULIRGs selected using {\it Spitzer} data have shown that they seem to resemble local ULIRGs; for example, \citet{saj07a} find that $z\gtrsim1$ ULIRGs selected on the basis of red 24$\mu$m/8$\mu$m and 24$\mu$m/0.7$\mu$m color often show PAHs accompanied by an AGN continuum and in some ($\sim25\%$) cases strong silicate absorption, and their figure 13 appears to place the bulk of their sample in the upper branch of Figure \ref{fork}. We might therefore expect that our sources should be distributed mainly along the upper branch in Figure \ref{fork}, along with the majority of local ULIRGs, and previous small samples of Spitzer selected high redshift ULIRGs. This however is not the case; our sources are located in a specific region, classifying them as starbursts with (relatively) weak apparent silicate absorption, and almost completely avoid the upper branch in Figure \ref{fork}, where most local ULIRGs reside. 

There is no {\it a priori} reason why our relatively crude selection method should select against ULIRGs with strong silicate absorption, and the almost complete absence of such absorption in our sample is peculiar, for two reasons. First, many local ULIRGs have a near-IR excess coupled with strong silicate absorption \citep{spi95,arm07}. Second, the 24$\mu$m flux cut (in combination with the demand for a rest-frame 1.6$\mu$m `bump') selects in favor of sources with strong PAHs, but among local ULIRGs, systems with strong PAHs but weak silicate absorption are rare - e.g. \citet{des07} examine ULIRG local spectra medianed by optical spectra type, and find strong PAHs and prominent silicate absorption in the HII and LINER composites, and weak PAHs and weak silicate absorption in the Seyfert composite. 

\begin{figure}
\begin{minipage}{180mm}
\includegraphics[angle=90,width=170mm]{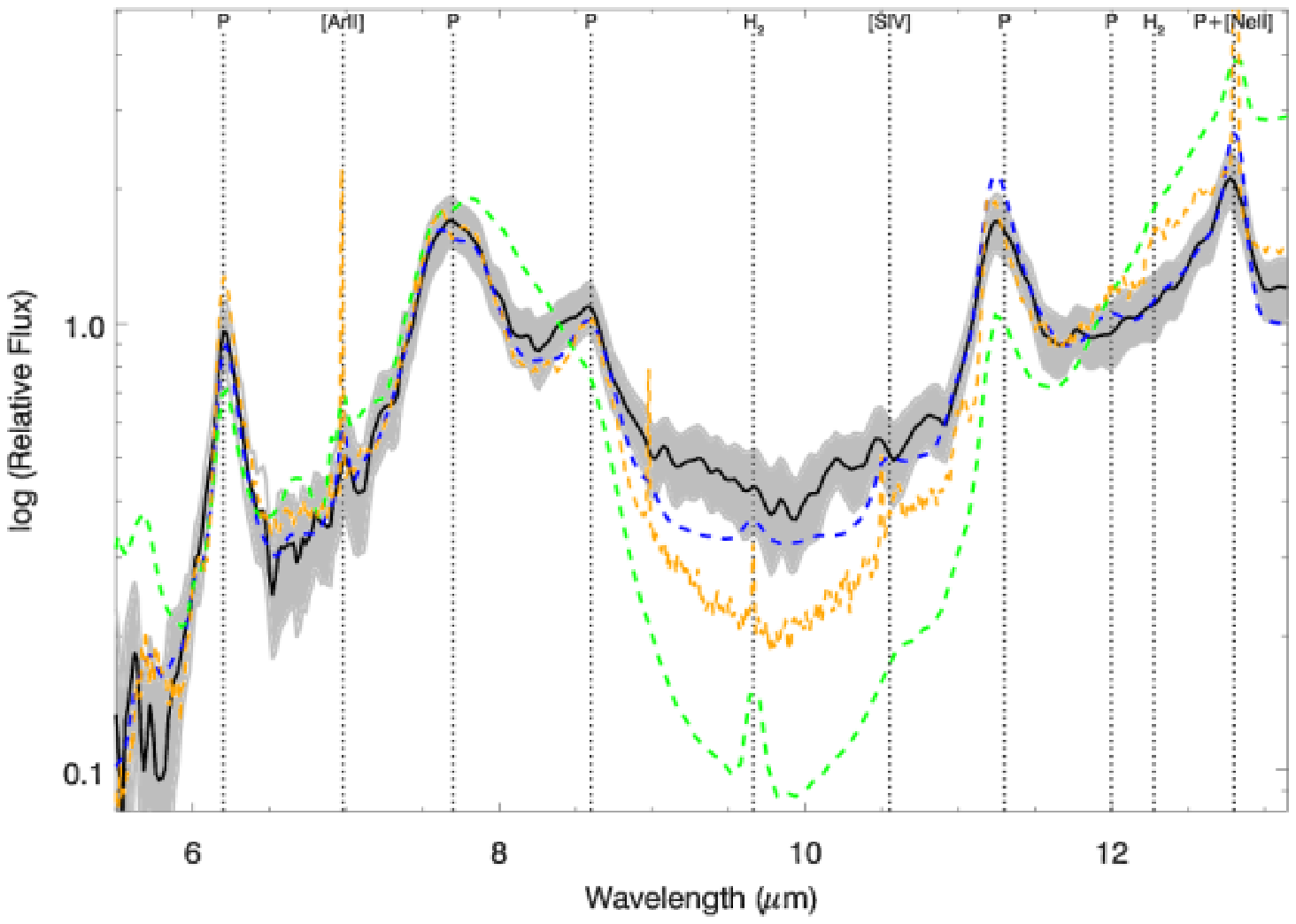}
\end{minipage}
\caption{Comparison between our averaged spectra (black, L$_{IR}\simeq$10$^{12.9}$ - 10$^{13.8}$L$_{\odot}$) and three different starburst galaxy templates, all normalized at 6$\mu$m; the averaged spectrum of the starburst galaxies in \citealt{bra06} (blue, L$_{IR}\simeq$10$^{10}$ - 10$^{11.5}$L$_{\odot}$), Arp 220 (\citealt{arm06}, green, L$_{IR}=$10$^{12.14}$L$_{\odot}$), and M82 (\citealt{stu00}, orange, L$_{IR}=$10$^{10}$L$_{\odot}$, this spectrum was taken with the ISO-SWS and hence is of higher spectral resolution than the other spectra). Despite being approximately two orders of magnitude less luminous, the Brandl et al spectrum is an excellent match to our average spectrum. M82 is also a reasonable match, except for slightly stronger apparent silicate absorption. Arp 220 on the other hand is a very poor match, with much stronger absorption. 
\label{comp_brandl}}
\end{figure}

If our sample do not resemble local ULIRGs, then what do they look like? We compare in Figure \ref{comp_brandl} our average spectrum with three different starburst templates; Arp 220 \citep{arm07}, the nucleus of M82 \citep{stu00}, and the average spectrum of the local starburst galaxies from \citealt{bra06} (which have IR luminosities between 10$^{10}$L$_{\odot}$ and 10$^{11.5}$L$_{\odot}$). Clearly, the Brandl et al spectrum is the closest match to our spectrum, even though the parent samples differ in IR luminosity by over two orders of magnitude. Both show strong PAHs with an asymmetric 12.7$\mu$m PAH profile, and negligible silicate absorption.  Interestingly, another class of high redshift galaxy that tends to show strong PAHs but weak silicate absorption are the sub-mm galaxies \citep{pop07,men07}. The M82 nuclear spectrum is also a good match, except for slightly stronger silicate absorption. The Arp 220 spectrum on the other hand is a poor match, with much stronger silicate absorption. 

The similarity of our composite spectrum to that of local starburst galaxies allows us to estimate the IR luminosities for our sample using the relation between IR luminosity and 6.2$\mu$m PAH luminosity in \citealt{bra06}:

\begin{equation}
log (L_{IR}) = 1.13\times log\left(F_{6.2}D_{L}^{2}\right) \label{pahlirs}
\end{equation}

\noindent where $F_{6.2}$ is the flux of the PAH 6.2$\mu$m feature in units of $10^{-19}$ W cm$^{-2}$, $D_{L}$ is the luminosity distance in Kpc, and $L_{IR}$ is in units of solar luminosities. The IR luminosities for our sample computed using Equation \ref{pahlirs} are presented in Table \ref{pahsils}, and are in good agreement with the IR luminosities estimated in \S\ref{takeaguess}. 

\begin{figure}
\begin{minipage}{180mm}
\includegraphics[angle=0,width=160mm]{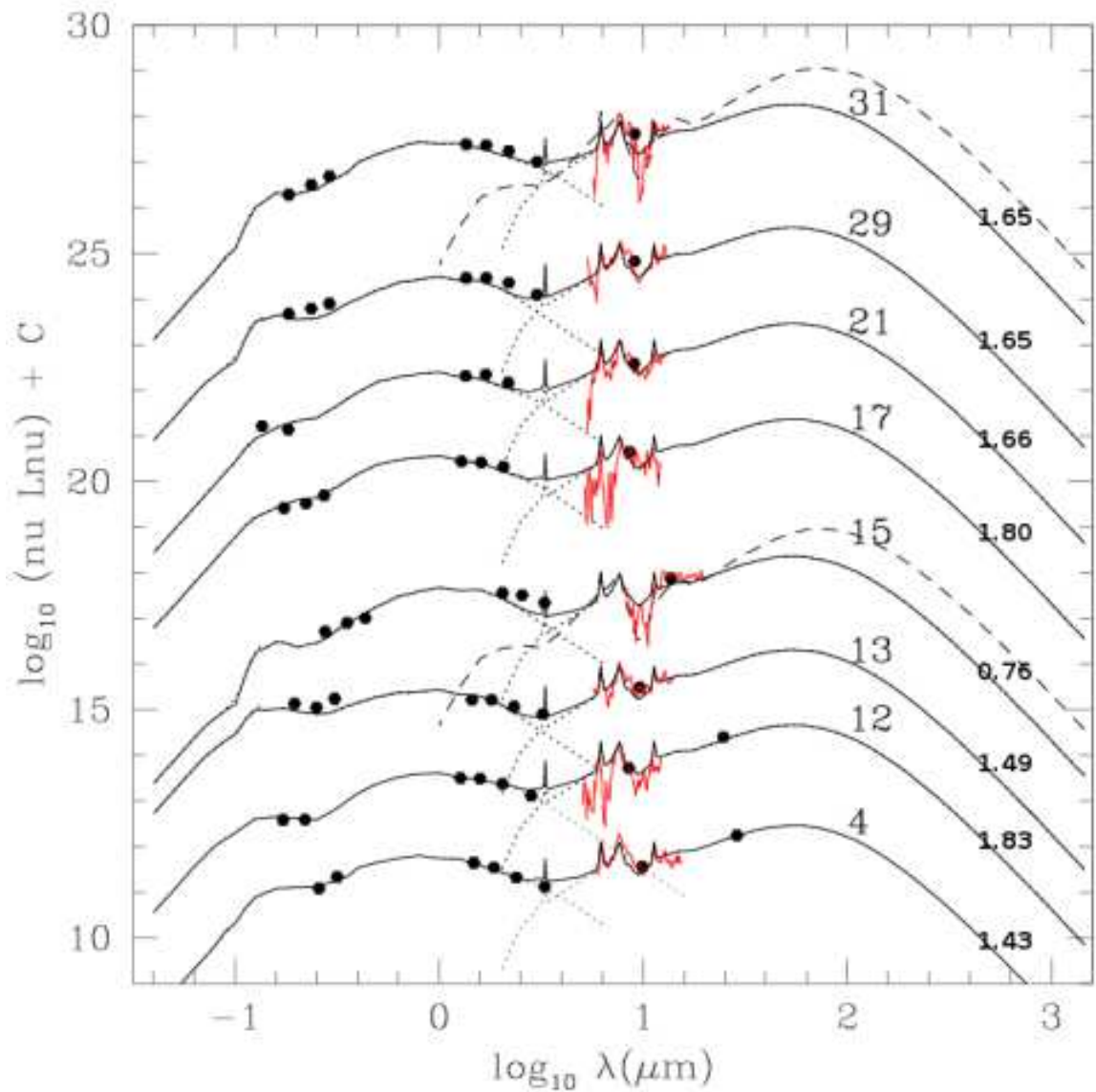}
\end{minipage}
\caption{Spectral energy distributions of the eight objects in our sample with at least six optical-IR photometric data points (black circles), plotted in the rest-frame. Overlaid are the best-fit templates derived from the \citet{rr07} code. Redshifts are fixed at the spectroscopic values, and are given on the right hand side. In each fit, the leftmost dotted line is the starlight component, the righmost dotted line is an M82 template, and the solid black line is the sum of the two. For two objects (15 \& 31) we show an alternative fit to the IR emission from an Arp 220 template (dashed line). The IRS spectra have been superimposed as red lines, but are not used in the fitting. 
\label{mrrdiag}}
\end{figure}

We explore the robustness of the conclusion that our sample have systematically weaker apparent silicate absorption than is seen in local ULIRGs using a different approach; fitting their broad band optical through IR SEDs (i.e. the $Ugriz$, IRAC and MIPS data) with the code of \citet{rr07}, with the redshifts fixed at the spectroscopic values, and comparing the results from these fits with the IRS spectra. We consider only those objects (4, 12, 13, 15, 17, 21, 29, 31) with detections in at least six optical through MIPS bands, and specifically do {\it not} fit to the IRS spectra, as we wish to test the conclusions independently. The limited available photometry (less than nine points in all cases) renders this approach crude compared to the analysis of the IRS spectra, but it serves as a useful check. The results from this fitting are shown in Figure \ref{mrrdiag}. In four cases (4, 13, 21 29), an M82 template gives an excellent fit to the photometry, and is well matched to the IRS spectrum. In two cases (12 and 17) an M82 template gives a good fit to the photometry, but predicts a higher flux at rest-frame 6$\mu$m than is seen in the IRS spectrum, which may indicate a steeper mid-IR continuum. In the final two cases (15 and 31) there is marginal evidence for strong silicate absorption, consistent with an Arp 220 template, though the low S/N of the IRS spectra means we cannot rule out an M82 template either. Overall, the results from the template fitting agree with the results from the individual IRS spectra, and from the comparison between our average spectrum and Arp 220, M82, and the \citealt{bra06} composite in Figure \ref{comp_brandl}.

All of the evidence points towards an unexpected result; despite the fact that our sample are ultraluminous in the IR, their mid-IR spectra have systematically lower silicate absorption than low-redshift ULIRGs, and instead resemble the mid-IR spectra of local starbursts that are over two orders of magnitude less luminous in the IR. What makes this even more puzzling is that high redshift ULIRGs appear to be physically similar to local ULIRGs in many ways; they show evidence for being mergers (\citealt{far02a,cha03,bri07}, though see also \citealt{gen06}) between massive, gas rich systems \citep{tac06,gen06}.

We can think of six possible reasons to explain this difference:

\noindent 1 - {\it IR Luminosity dependent silicate strength:} Our sample is significantly more IR-luminous than most local ULIRGs, so we would see weak apparent silicate absorption if compact, dusty starbursts and AGN exhibit weaker silicate absorption as they increase in IR luminosity, but remain otherwise identical. While we cannot rule this possibility out, we can think of no plausible physical reason why this should be the case. 

\noindent 2 - {\it A specific evolutionary stage:} We may be selecting high redshift ULIRGs at a specific evolutionary stage, when a compact but relatively old starburst ($\gtrsim20$Myr if we assume an exponentially decaying burst) becomes less obscured as supernova winds clear dust from the star-forming regions. While this is in principle plausible, our selection method is relatively crude compared to others cited as selecting distant ULIRGs at specific evolutionary stages, which rely on the presence of specific fine-structure absorption lines \citep{far07a}. It seems unlikely that our method should find all 32 of our sample at this specific stage. 

\noindent 3 - {\it Silicate poor dust:} Weak apparent silicate absorption could arise from dust that is unusually poor in silicates. If the dust is made mainly in Asymptotic Giant Branch (AGB) stars \citep{geh89}, then silicate poor dust could be made in two ways; a low metallicity (Z$\lesssim0.04$) would suppress the formation of silicate dust, and a steep or `top-heavy' stellar IMF may shift more AGB stars into the carbon dust producing mass range, thereby reducing the production of silicate dust \citep{bla78,ren81,lag07}. The metallicities of distant ULIRGs are not well constrained \citep{sha04,lia04}, but it seems unlikely that the metallicities of our sample could be lowered to the extent required, as low metallicities also suppress PAH features \citep{eng05,smi07,wu07}. A top heavy IMF is more plausible, and is consistent with the demands from structure formation models (e.g. \citealt{bau}). Interestingly, if the IMF were top heavy, it would result in a harsher radiation field, which (as described in \S\ref{pahfluxsect}) may result in enhanced 6.2$\mu$m PAH emission compared to the longer wavelength PAH features, a result hinted at in Figures \ref{pahfluxratio} and \ref{comp_brandl}. There is however no direct evidence for a top-heavy IMF in distant ULIRGs. Moreover, some authors have suggested there is no evidence for evolution in dust properties with redshift, at least up to $z\sim1$ (e.g. \citealt{cle05,bav07}).

\noindent 4 - {\it An intrinsically weak underlying continuum:} If the underlying near/mid-IR continua in our sample are weak, then we would see three effects; the rest-frame 1.6$\mu$m feature would be prominent, the PAH EWs would be higher, and we would see weak silicate absorption as there would be no background continuum upon which to make such absorption apparent. A weak continuum would arise if our sample harbored lower luminosity AGN than those seen in local ULIRGs. In this scenario the absence of silicate absorption, and the differences between our sample and that of \citet{saj07a}, arises due to selection bias, as our selection method favors systems harboring active star formation. It is also possible that the differences between our sample and that of \citet{saj07a} arise because our sample is fainter at observed-frame 24$\mu$m (most of our sample fall below the 24$\mu$m flux cut of 0.9mJy used by \citet{saj07a}) which, if true, would imply a `transition' in the mid-IR properties of high redshift ULIRGs at $f_{24}\lesssim1$mJy. This possibility is simple, but does suffer from three drawbacks. First, as described in \S\ref{select}, we do not demand that the 1.6$\mu$m bump is detected with any statistical significance, and many of our sources have a weak 1.6$\mu$m bump. Our selection method therefore does not demand that the underlying continuum in the rest-frame near-IR is negligible, and therefore by extension does not demand that the rest-frame mid-IR continuum is absent either. Second, as can be seen in Figures \ref{spectraa} and \ref{spectrab}, several of our sample have significant mid-IR continua. Third, several local ULIRGs with strong mid-IR continua, in some cases accompanied by silicate absorption, also show a near-IR `bump' (e.g. Arp 220, NGC6240, \citealt{spi95}), hence the presence of a bump does not necessarily select against objects in the upper branch of Figure \ref{fork}.

\noindent 5 - {\it Gas-rich mergers:} A simple way to reduce the apparent silicate strength is to lower the total extinction, which can be achieved by reducing the amount of dust in the starburst regions, or equivalently by increasing the amount of gas so that the same level of star formation can be sustained for a smaller dust mass. This scenario is compatible with the presence of strong PAH emission, as changing the dust-to-gas ratio mainly affects the total dust column to the central ionizing source, whereas PAH emission is thought to originate from the surface layers of star-forming regions. There is some support for this possibility from millimeter interferometry of sub-mm selected ULIRGs at high redshift, which shows that they may be unusually gas-rich compared to local examples \citep{tac06}, and also from models for the star formation history of the Universe, which suggest a lower mean $A_{V}$ at $z\sim1.5$ compared to locally (e.g. \citealt{rr03}, their figure 20).

\noindent 6 - {\it Starburst geometry:} The final possibility is a different star-forming region geometry, such that the total dust column is reduced. This can be achieved in two ways; either make the star formation diffuse and extended over several Kpc instead of the sub-Kpc scale starbursts seen in local ULIRGs\footnote{Such extended star formation would however still give rise to PAH emission, see for example the `cirrus' models of \citealt{efs03}}, or by distributing the star formation in multiple compact dusty star forming regions spread over a few Kpc, instead of a single sub-Kpc nuclear starburst. There is observational evidence for both these `modes' of star formation in distant starburst galaxies; there is evidence for extended star formation from high resolution radio observations \citep{cha04,mux05} and Ly$\alpha$ imaging \citep{gea05}, while other authors have found evidence for star formation in multiple small regions distributed across a few Kpc in forming massive galaxies at high redshift \citep{gen06}. Furthermore, \citet{dad07} have recently suggested that disklike, `inefficient' star formation may be prevalent among distant massive galaxies. There is however an observational limit on the spatial extents of these starbursts; high resolution millimeter imaging of small samples of sub-mm selected ULIRGs at high redshift have shown that the starbursts in these systems cannot be much more than 4Kpc in diameter \citep{tac06}. 

It is difficult to discriminate reliably between these six possibilities, as the IRS spectra cannot formally rule any of them out. The first two possibilities however do not seem likely; one would require some as-yet unknown dependence on luminosity for the silicate feature, while the other would require an extraordinary level of serendipity to find nearly all of the sample in the same evolutionary stage. The third possibility is more plausible, but would require a `top-heavy' IMF, for which there is currently no direct evidence. The fourth possibility is also plausble, and some level of selection bias is undoubtedly present in our sample as we are not selecting solely on the basis of IR luminosity, but selection bias likely plays only a minor role, for the reasons given above.  We tentatively propose therefore that some combination of the final two possibilities is the most likely origin for most (but not all) of this effect. It is worth noting that these conclusions may also apply to sub-mm selected ULIRGs, given the similarity between their respective IRS spectra.

\section{Conclusions}\label{conc}
We have presented mid-infrared spectra, taken using the IRS onboard Spitzer, of 32 objects selected on (1) an optical band magnitude fainter than $m_{r}=23$, (2) excess flux in the IRAC 4.5$\mu$m passband compared to the other three IRAC channels, and (3) a 24$\mu$m flux in excess of 0.5mJy. Our conclusions are:

\noindent 1 - All of the sample lie within $0.8<z<2.3$, with more than 90\% lying in a much tighter redshift range, characterized by $\langle z \rangle=1.71\pm0.15$. The rest-frame IR luminosities are consistent with the range $10^{12.9}-10^{13.8}$L$_{\odot}$. Our selection method is therefore a straightforward and efficient way of finding large samples of distant ULIRGs confined to narrow redshift ranges. It avoids all of the observational challenges inherent in finding distant ULIRGs via their far-IR emission, and produces samples with redshift distributions akin to those of optically selected samples such as Lyman-Break galaxies. 

\noindent 2 - Nearly all of the sample show two or more strong PAH emission features in their mid-IR spectra, and systematically lack strong silicate absorption. The individual spectra closely resemble the IRS spectra of high-redshift sub-mm selected ULIRGs. The star formation rates computed from the PAH luminosities are all extremely high, of order $\gtrsim1000$M$_{\odot}$yr$^{-1}$, sufficient to power all of the IR emission. Coupled with the lack of a power law continuum in the IRAC channels, we infer that star formation is likely to be the dominant power source in most of the sample, though it is possible that some also contain a luminous AGN. 

\noindent 3 - Our sample occupy a different region on the PAH 6.2$\mu$m EW vs silicate strength plane to most local ULIRGs, and there is marginal evidence for enhanced PAH 6.2$\mu$m/7.7$\mu$m and 6.2$\mu$m/11.2$\mu$m ratios. Furthermore, the composite mid-IR spectrum of our sample resembles that of local starburst galaxies with IR luminosities of $10^{10}-10^{11.5}$L$_{\odot}$, rather than that of local ULIRGs. Though selection bias likely plays a role, we propose that the most likely reasons for this are that the star formation in distant ULIRGs is extended over scales of a few Kpc rather than the sub-Kpc starbursts seen in local ULIRGs, and/or occurs in unusually gas-rich environments.

\acknowledgments
We thank the referee for a very helpful report. This work is based on observations made with the Spitzer Space Telescope, which is operated by the Jet Propulsion Laboratory, California Institute of Technology under a contract with NASA. Support for this work by the IRS GTO team at Cornell University was provided by NASA through Contract Number 1257184 issued by JPL/Caltech. Support for this work was provided by NASA. This research has made extensive use of the NASA/IPAC Extragalactic Database (NED) which is operated by the Jet Propulsion Laboratory, California Institute of Technology, under contract with NASA. MP acknowledges financial support from the Marie-Curie Fellowship grant MEIF-CT-2007-042111

\clearpage

\begin{deluxetable}{ccclcccccccccc}
\tabletypesize{\scriptsize}
\tablecolumns{14}
\tablewidth{0pc}
\tablecaption{Observations summary \label{sample}}
\tablehead{
\colhead{ID}&\colhead{Galaxy}&\colhead{AOR key}&\colhead{RA (J2000)}&\colhead{Dec}&\colhead{3.6$\mu$m}&\colhead{4.5$\mu$m}&\colhead{5.8$\mu$m}&\colhead{8$\mu$m}&\colhead{24$\mu$m}
}
\startdata
1  & SWIRE3 J103205.16+574817.5 & 17420288 & 158.02150 &  57.804861  & 68  & 87  & 74  & 75  & 699  \\
2  & SWIRE3 J103707.79+591204.5 & 17422080 & 159.28250 &  59.201250  & 42  & 50  & 50  & 37  & 625  \\
3  & SWIRE3 J104011.60+580542.6 & 17416704 & 160.04837 &  58.095194  & 63  & 76  & 65  & --  & 751  \\
4  & SWIRE3 J104012.86+592712.4 & 17414912 & 160.05358 &  59.453472  & 150 & 150 & 116 & 101 & 838  \\ 
5  & SWIRE3 J104034.35+582314.5 & 17417472 & 160.14312 &  58.387389  & 42  & 49  & 54  & --  & 680  \\%
6  & SWIRE3 J104129.25+581712.2 & 17421824 & 160.37192 &  58.286750  & 43  & 62  & 46  & --  & 561  \\
7  & SWIRE3 J104139.78+573723.8 & 17421056 & 160.41575 &  57.623306  & 40  & 52  & 48  & --  & 601  \\
8  & SWIRE3 J104232.04+575439.4 & 17418240 & 160.63350 &  57.910972  & 37  & 49  & --  & --  & 704  \\
9  & SWIRE3 J104343.93+571322.5 & 17414656 & 160.93304 &  57.222917  & 97  & 116 & 112 & 69  & 1009 \\
10 & SWIRE3 J104349.41+575438.7 & 17419776 & 160.95592 &  57.910750  & 78  & 102 & 72  & 62  & 700  \\
11 & SWIRE3 J104402.55+593204.6 & 17421568 & 161.01067 &  59.534639  & 63  & 73  & 77  & --  & 615  \\
12 & SWIRE3 J104427.54+593811.7 & 17418752 & 161.11475 &  59.636611  & 62  & 75  & 76  & 57  & 686  \\
13 & SWIRE3 J104436.55+593252.4 & 17416448 & 161.15233 &  59.547889  & 51  & 64  & 58  & 55  & 644  \\
14 & SWIRE3 J104514.38+575708.8 & 17415424 & 161.30996 &  57.952472  & 65  & 78  & 70  & --  & 874  \\
15 & SWIRE3 J104551.62+594234.8 & 17416192 & 161.46513 &  59.709694  & 49  & 55  & 49  & --  & 703  \\
16 & SWIRE3 J104614.90+594134.3 & 17420800 & 161.56208 &  59.692861  & 42  & 52  & --  & --  & 715  \\
17 & SWIRE3 J104627.82+592843.4 & 17420544 & 161.61596 &  59.478722  & 56  & 67  & 69  & --  & 607  \\
18 & SWIRE3 J104632.93+563530.2 & 17416960 & 161.63721 &  56.591750  & 76  & 91  & 68  & --  & 761  \\
19 & SWIRE3 J104643.29+575851.0 & 17415936 & 161.68037 &  57.980833  & 82  & 98  & 66  & 64  & 772  \\
20 & SWIRE3 J104653.07+592652.1 & 17421312 & 161.72117 &  59.447833  & 86  & 100 & 87  & --  & 637  \\
21 & SWIRE3 J104656.23+594008.0 & 17419264 & 161.73433 &  59.668889  & 50  & 67  & 57  & --  & 632  \\
22 & SWIRE3 J104754.66+583905.9 & 17418496 & 161.97775 &  58.651667  & 48  & 60  & 38  & --  & 688  \\
23 & SWIRE3 J104843.21+584537.8 & 17417728 & 162.18008 &  58.760500  & 48  & 62  & 48  & --  & 687  \\
24 & SWIRE3 J104845.19+561055.8 & 17419008 & 162.18833 &  56.182194  & 69  & 75  & --  & --  & 674  \\
25 & SWIRE3 J104922.64+564032.5 & 17417984 & 162.34438 &  56.675694  & 93  & 110 & 103 & 71  & 645  \\
26 & SWIRE3 J105056.08+562823.0 & 17419520 & 162.73371 &  56.473056  & 69  & 80  & 67  & 52  & 621  \\
27 & SWIRE3 J105152.73+564719.2 & 17417216 & 162.96975 &  56.788667  & 65  & 76  & 59  & --  & 590  \\
28 & SWIRE3 J105308.24+591447.5 & 17415680 & 163.28433 &  59.246528  & 75  & 88  & 79  & --  & 793  \\
29 & SWIRE3 J105334.59+574242.3 & 17414400 & 163.39417 &  57.711750  & 71  & 128 & 91  & 69  & 1108 \\
30 & SWIRE3 J105539.92+571711.8 & 17415168 & 163.91637 &  57.286639  & 74  & 100 & 85  & 66  & 863  \\
31 & SWIRE3 J105908.46+574511.4 & 17420032 & 164.78525 &  57.753167  & 60  & 74  & 71  & 56  & 688  \\
32 & SWIRE2 J161744.64+540031.4 & 16161792 & 244.43583 &  54.008722  & 62  & 88  & 68  & 69  & 1074 \\
\enddata  
\tablecomments{All fluxes are given in $\mu$Jy. Errors on the IRAC fluxes are of order $5\%$, while the errors on the MIPS 24$\mu$m fluxes are of order 10$\%$.  A `--' indicates an upper limit. Limiting fluxes are given in \S\ref{select}.}

\end{deluxetable}

\clearpage

\begin{deluxetable}{clcc}
\tabletypesize{\scriptsize}
\tablecolumns{4}
\tablewidth{0pc}
\tablecaption{Redshifts \& bump classifications \label{measurements}}
\tablehead{
 \colhead{ID}& \colhead{Bump\tablenotemark{a}}&\colhead{$z_{phot}$\tablenotemark{b}}&\colhead{$z_{irs}$\tablenotemark{c}}
}
\startdata 
1  & 2   &  --     & 1.63  \\   
2  & 2/3 &  1.88?  & 1.84  \\   
3  & 2   &  --     & 1.83 \\    
4  & 1/2 &  1.18   & 1.43  \\ 
5  & 2/3 &  1.73   & 1.89  \\    
6  & 2   &  --     & 1.61  \\   
7  & 2   &  --     & 1.69  \\   
8  & 2   &  --     & 1.93  \\  
9  & 2   &  --     & 1.71  \\ 
10 & 2   &  --     & 1.77  \\ 
11 & 2/3 &  1.00?  & 1.74?  \\ 
12 & 2/3 &  2.03   & 1.83  \\ 
13 & 2   &  1.61   & 1.49 \\     
14 & 2   &  --     & 1.78  \\ 
15 & 2   &  1.30   & 0.76? \\ 
16 & 2   &  0.58   & 2.11  \\
17 & 2/3 &  1.63   & 1.80  \\   
18 & 2   &  --     & 1.76 \\  
19 & 2   &  2.36?  & 1.60  \\   
20 & 2   &  --     & 1.69  \\  
21 & 2   &  1.61   & 1.66  \\        
22 & 2   &  --     & 1.63  \\  
23 & 2   &  --     & 2.23  \\     
24 & 2   &  --     & 1.25  \\
25 & 2   &  --     & 1.70  \\ 
26 & 2   &  --     & 1.54  \\  
27 & 2   &  --     & 1.88?  \\
28 & 2   &  --     & 1.47  \\  
29 & 2   &  1.55   & 1.65 \\ 
30 & 2   &  --     & 1.68  \\     
31 & 2/3 &  1.42   & 1.65  \\ 
32 & 2   &  1.81?  & 1.80  \\   
\enddata  
\tablenotetext{a}{Bump classification, based on the IRAC fluxes in Table \ref{sample}}
\tablenotetext{b}{Photometric redshifts taken from \citet{rr07}. The objects with a `?' are based on three photometric bands, and should be treated with caution.}
\tablenotetext{c}{Spectroscopic redshifts derived from the IRS spectra. The three measurements with a `?' should be treated with caution.}
\end{deluxetable}

\begin{deluxetable}{cccccccccc}
\tabletypesize{\scriptsize}
\tablecolumns{10}
\tablewidth{0pc}
\tablecaption{Fluxes and equivalent widths for major PAH features, silicate strengths, and star formation rates\label{pahsils}}
\tablehead{
\colhead{ID}&\multicolumn{2}{c}{PAH 6.2$\mu$m}&\multicolumn{2}{c}{PAH 7.7$\mu$m}&\multicolumn{2}{c}{PAH 11.2$\mu$m}&\colhead{$S_{sil}$} &\colhead{SFR\tablenotemark{a}} &\colhead{L$_{IR}$\tablenotemark{b}}\\
\colhead{  }&\colhead{Flux}&\colhead{EW}&\colhead{Flux}&\colhead{EW}&\colhead{Flux}&\colhead{EW}&\colhead{}&\colhead{M$_{\odot}$ yr$^{-1}$}&\colhead{log(L$_{\odot}$)}
}
\startdata  
1  &  3.88 $\pm$ 1.13 & 1.01 $\pm$  1.14 &  5.67 $\pm$ 1.53 & 0.53 $\pm$  0.26 &  2.68 $\pm$  0.58 &  0.63 $\pm$  0.20 &  0.00 $\pm$  0.45 &    1352 & 13.28 \\      
2  &  $<$4.13         & $<$0.70          &  $<$5.82         & 0.43 $\pm$  0.31 &  2.53 $\pm$  0.70 &  0.48 $\pm$  0.19 &  0.35 $\pm$  0.53 & $<$1850 & $<$13.46 \\ 
3  &  $<$4.49         & $<$5.39          &  8.62 $\pm$ 1.97 & 0.77 $\pm$  0.28 &  3.99 $\pm$  0.95 &  1.44 $\pm$  1.01 &  1.07 $\pm$  1.00 &    2324 & 13.49 \\       
4  &  8.82 $\pm$ 0.99 & 0.64 $\pm$  0.21 & 17.79 $\pm$ 1.40 & 0.61 $\pm$  0.10 &  5.03 $\pm$  0.45 &  0.51 $\pm$  0.07 &  0.17 $\pm$  0.15 &    2067 & 13.53 \\       
5  &  2.47 $\pm$ 0.95 & 2.66 $\pm$ 14.15 &  5.27 $\pm$ 1.35 & 0.58 $\pm$  0.23 &  1.55 $\pm$  0.77 &  0.37 $\pm$  0.32 &  0.24 $\pm$  0.62 &    1194 & 13.24 \\       
6  &  3.15 $\pm$ 1.12 & 0.52 $\pm$  0.43 &  9.22 $\pm$ 1.82 & 0.80 $\pm$  0.47 &  2.39 $\pm$  0.44 &  0.50 $\pm$  0.16 &  0.68 $\pm$  0.37 &    1108 & 13.16 \\        
7  &  $<$4.40         & $<$3.16          &  5.96 $\pm$ 1.46 & 0.69 $\pm$  0.36 &  1.72 $\pm$  0.46 &  0.39 $\pm$  0.14 &  0.06 $\pm$  0.43 & $<$1378 & $<$13.39 \\       
8  &  4.18 $\pm$ 1.36 & 0.91 $\pm$  1.04 &  3.98 $\pm$ 0.79 & 0.34 $\pm$  0.09 &  2.65 $\pm$  0.70 &  0.52 $\pm$  0.20 &  0.87 $\pm$  0.39 & 2133 & 13.52 \\          
9  &  7.49 $\pm$ 0.95 & 1.88 $\pm$  2.14 & 11.82 $\pm$ 1.71 & 0.79 $\pm$  0.19 &  4.86 $\pm$  0.63 &  0.62 $\pm$  0.13 &  0.55 $\pm$  0.25 & 1160 & 13.27 \\           
10 &  $<$4.35         & $<$0.64          &  7.52 $\pm$ 1.85 & 0.77 $\pm$  0.32 &  2.95 $\pm$  0.61 &  1.38 $\pm$  1.24 &  1.19 $\pm$  0.80 & $<$1844 & $<$13.44 \\    
11 &  $<$4.03         & $<$0.83          &  $<$6.0          & $<$4.00          &  $<$1.79          & $<$0.34           &  0.02 $\pm$  0.40 & $<$1410 & $<$13.38 \\    
12 &  3.46 $\pm$ 1.05 & 1.02 $\pm$  1.39 &  4.48 $\pm$ 0.84 & 0.38 $\pm$  0.10 &  $<$1.86          & $<$0.41           &  0.67 $\pm$  0.43 & $<$1458 & 13.37 \\       
13 &  4.72 $\pm$ 1.38 & 0.63 $\pm$  0.59 &  4.62 $\pm$ 1.53 & 0.22 $\pm$  0.08 &  2.30 $\pm$  0.42 &  0.25 $\pm$  0.06 &  0.20 $\pm$  0.26 &  1160 & 13.27 \\        
14 &  4.91 $\pm$ 0.96 & 1.26 $\pm$  1.52 &  6.84 $\pm$ 1.51 & 0.50 $\pm$  0.17 &  2.87 $\pm$  0.62 &  0.66 $\pm$  0.29 &  0.67 $\pm$  0.62 & 1991 & 13.50 \\        
15 &  --              & --               &  --              & --               &  $<$4.37          &  $<$1.49          &   1.89 $\pm$  0.43\tablenotemark{c} & -- & --\\
16 &  $<$5.07         & $<$1.23          &  2.80 $\pm$ 0.93 & 0.22 $\pm$  0.08 &   --              &  --               &  --                & -- & $<$13.72 \\        
17 &  1.53 $\pm$ 0.54 & 0.63 $\pm$  1.09 &  3.94 $\pm$ 1.84 & 0.57 $\pm$  0.33 &  2.33 $\pm$  0.00 &  0.59 $\pm$  0.00 &  0.01 $\pm$  0.44 &  1014 & 12.94 \\        
18 &  3.96 $\pm$ 1.49 & 0.98 $\pm$  1.01 &  4.25 $\pm$ 1.36 & 0.26 $\pm$  0.09 &  2.89 $\pm$  0.69 &  0.53 $\pm$  0.19 &  0.45 $\pm$  0.43 &  1706 & 13.38 \\        
19 &  4.94 $\pm$ 1.06 & 0.95 $\pm$  0.75 &  9.65 $\pm$ 1.98 & 0.59 $\pm$  0.18 &  2.99 $\pm$  0.65 &  0.44 $\pm$  0.14 &  0.27 $\pm$  0.51 &  1562 & 13.38 \\        
20 &  $<$4.16         & $<$0.81          &  $<$6.00         & $<$1.50           &  4.23 $\pm$  0.52 &  0.93 $\pm$  0.24 &  0.40 $\pm$  0.45 & $<$3008 & $<$13.74 \\   
21 &  4.35 $\pm$ 0.98 & 0.56 $\pm$  0.35 &  7.64 $\pm$ 1.42 & 0.54 $\pm$  0.18 &  1.45 $\pm$  0.38 &  0.24 $\pm$  0.08 &  0.58 $\pm$  0.31 &  1423 & 13.42 \\         
22 &  5.85 $\pm$ 1.10 & 7.34 $\pm$ 40.91 &  8.28 $\pm$ 1.59 & 0.58 $\pm$  0.21 &  3.89 $\pm$  0.66 &  0.71 $\pm$  0.23 &  0.54 $\pm$  0.52 &  2009 & 13.48 \\         
23 &  $<$4.86         & $<$0.49          &  4.68 $\pm$ 0.76 & 0.51 $\pm$  0.15 &  --               &  --               &  --               & -- & $<$13.77 \\          
24 &  --              & --               &  8.16 $\pm$ 1.43 & 0.31 $\pm$  0.07 &  8.83 $\pm$  1.29 &  3.32 $\pm$  3.05 &  0.83 $\pm$  0.64 &   --   & --    \\
25 &  $<$3.10         & $<$0.28          &  $<$5.00         & 0.25 $\pm$  0.11 &  2.39 $\pm$  0.68 &  1.06 $\pm$  0.66 &  1.34 $\pm$  1.34 & $<$1253 & $<$13.22 \\      
26 &  8.39 $\pm$ 1.21 & 0.65 $\pm$  0.22 &  7.22 $\pm$ 1.47 & 0.28 $\pm$  0.07 &  5.13 $\pm$  0.75 &  0.66 $\pm$  0.15 &  0.51 $\pm$  0.35 &  2424 & 13.59 \\         
27 &  $<$3.31         & $<$0.65          &  $<$4.80         & 0.41 $\pm$  0.22 &  $<$1.30          &  $<$0.26          &  0.01 $\pm$  0.43 & $<$1281 & $<$13.35 \\     
28 &  3.49 $\pm$ 1.19 & 0.47 $\pm$  0.45 &  5.97 $\pm$ 1.23 & 0.31 $\pm$  0.09 &  2.96 $\pm$  0.40 &  0.35 $\pm$  0.06 &  0.00 $\pm$  0.24 & 1030 & 13.10 \\          
29 &  6.22 $\pm$ 1.10 & 0.59 $\pm$  0.25 & 10.56 $\pm$ 1.57 & 0.45 $\pm$  0.09 &  2.86 $\pm$  0.55 &  0.24 $\pm$  0.05 &  0.73 $\pm$  0.27 &  1929 & 13.53 \\          
30 &  4.50 $\pm$ 1.12 & 1.08 $\pm$  1.06 &  5.77 $\pm$ 1.62 & 0.33 $\pm$  0.11 &  2.40 $\pm$  0.50 &  0.33 $\pm$  0.09 &  0.37 $\pm$  0.32 & 1533 & 13.39 \\       
31 &  4.81 $\pm$ 0.93 & 2.62 $\pm$  5.95 &  9.36 $\pm$ 1.18 & 0.77 $\pm$  0.31 &  1.64 $\pm$  0.49 &  0.20 $\pm$  0.07 &  0.85 $\pm$  0.47 & 1370 & 13.40 \\       
32 &  5.80 $\pm$ 0.90 & 0.36 $\pm$  0.10 &  4.40 $\pm$ 0.96 & 0.26 $\pm$  0.07 &  $<$1.85          &  $<$0.19          &  0.02 $\pm$  0.50 &  $<$2017 & 13.60 \\       
\enddata  

\tablecomments{Fluxes are given in units of $10^{-22}$W cm$^{-2}$ and equivalent widths are given in $\mu$m. A `--' indicates the feature lies outside the bandpass.}
\tablenotetext{a}{Star formation rates for those sources where both the 6.2$\mu$m and 11.2$\mu$m PAH features lie within the IRS bandpass. The errors on all values are of order $50\%$, and assume that the star formation is similar in nature to that seen in local ULIRGs, but see \S \ref{nature}}
\tablenotetext{b}{Rest-frame IR luminosity, computed using Equation \ref{pahlirs}}
\tablenotetext{c}{Measurement is inherently uncertain as the feature is not reliably identified, and only half in the bandpass}

\end{deluxetable}

\end{document}